\documentclass[10pt,journal,compsoc]{IEEEtran}

\usepackage{tabularx}
    \newcolumntype{L}{>{\raggedright\arraybackslash}X}
\usepackage{comment}
\usepackage{amsmath}
\DeclareMathOperator*{\argmax}{argmax}
\usepackage{amssymb}
\usepackage{gensymb}
\usepackage{graphicx}
\usepackage{epstopdf} % Place after graphicx package.
\usepackage{color}
\usepackage{enumerate}
\usepackage{multirow}
\usepackage{cite}
\usepackage{multicol}
\usepackage{threeparttable}
\usepackage[ruled]{algorithm}
\usepackage[noend]{algorithmic}
\usepackage{subfig}
\usepackage{float}
\usepackage{mathtools}
\usepackage{array}
\newcolumntype{P}[1]{>{\centering\arraybackslash}p{#1}}
\newcolumntype{M}[1]{>{\centering\arraybackslash}m{#1}}
\usepackage{nopageno}
\usepackage{pifont}
\newcommand{\cmark}{\ding{51}}
\newcommand{\xmark}{\ding{55}}

\usepackage{url}

\usepackage[numbers]{natbib} 
\usepackage{xcolor}

\begin{document}

\title{Machine Learning Assisted Security Analysis of 5G-Network-Connected Systems}

\author{Tanujay Saha, Najwa Aaraj, and Niraj K. Jha~\IEEEmembership{(Fellow,~IEEE)}

\IEEEcompsocitemizethanks{\IEEEcompsocthanksitem 
This work was supported by NSF under Grant No. CNS-1617628.  T. Saha, and N. K. Jha are with 
the Department of Electrical Engineering, Princeton University, New Jersey,
NJ, 08544 (\{tsaha,jha\}@princeton.edu.  N. Aaraj is with Technology Innovation Institute, UAE 
(najwa@tii.ae).
}
}

\IEEEtitleabstractindextext{
\begin{abstract}
The core network architecture of telecommunication systems has undergone a paradigm shift in the 
fifth-generation (5G) networks. 5G networks have transitioned to software-defined 
infrastructures, thereby reducing their dependence on hardware-based network functions. New 
technologies, like network function virtualization and software-defined networking, have been 
incorporated in the 5G core network (5GCN) architecture to enable this transition. This has 
resulted in significant improvements in efficiency, performance, and robustness of the networks. 
However, this has also made the core network more vulnerable, as software systems are generally 
easier to compromise than hardware systems. In this article, we present a comprehensive 
security analysis framework for the 5GCN. The novelty of this approach lies in the creation and 
analysis of attack graphs of the software-defined and virtualized 5GCN through machine learning. 
This analysis points to 119 novel possible exploits in the 5GCN. We demonstrate that these possible exploits of 
5GCN vulnerabilities generate five novel attacks on the 5G Authentication and Key Agreement 
protocol. We combine the attacks at the network, protocol, and the application layers to 
generate complex attack vectors. In a case study, we use these attack vectors to find four novel 
security loopholes in WhatsApp running on a 5G network.
\end{abstract}

}

\maketitle
\IEEEdisplaynontitleabstractindextext
\IEEEpeerreviewmaketitle

\IEEEraisesectionheading{\section{Introduction}}
\label{Intro}
\IEEEPARstart{F}{ifth}-generation (5G) networks hold promise for realizing the vision of 
universal connectivity. They enable various verticals like Internet-of-Things (IoT), autonomous 
vehicles, smart cities, and telemedicine. These applications require high-bandwidth, robust, 
flexible, dynamic, and fault-tolerant network architectures. 

5G networks represent a huge leap, both qualitatively and quantitatively, from 
previous-generation telecommunication networks. The network core architecture has undergone a 
paradigm shift from its predecessor, the Evolved Packet Core. Previously, network functions were 
implemented on commodity hardware. In 5G networks, the network functions are mostly implemented 
in software. Moreover, with the advent of cloud computing, many network operations are now 
virtualized. This allows multiple operators to use the same underlying hardware resources to 
provide network services. This technology is broadly known as network function virtualization 
(NFV). 5G networks also separate the communication on the data plane from that on the control 
plane. This involves the use of a controller that observes the entire network before making 
routing decisions. This technology, broadly referred to as software-defined networking (SDN), 
has been shown to reduce both the operational expenditure (OPEX) and capital expenditure (CAPEX) 
of the network. Many of these transitions have become possible due to the utilization of an 
mm-wave technology in 5G. Incorporation of these new technologies results in 
significant improvements in efficiency, reliability, and flexibility of wireless networks. 
%\cite{condoluci2015softwarization, ordonez2017network}. 

The confluence of the new technologies makes the 5G core network (5GCN) an intricate system 
comprising SDN, NFV, distributed systems, and cloud computing. 
%\cite{panwar2016survey}. 
The 5GCN has a service-based architecture that dynamically modifies itself according to the 
requirements of the operators and users. However, introduction of new technologies into the 5GCN also 
expands its attack surface \cite{sens2018ready}, as it now inherits the vulnerabilities of all 
these individual technologies.

Prior work in 5G security has referred to broad categories of attacks that the 5GCN may be 
vulnerable to \cite{jover2019security, ahmad2018overview, sriram20195g}. In this article, we 
address far-reaching implications of these threats and how they may interact with each other to 
give rise to complex attacks that were infeasible in previous generations of telecommunication 
networks. The sequences of operations that are executed to implement an attack, also referred to 
as attack vectors, can be combined into an attack graph for a concise representation.  We 
combine the various attack vectors pertaining to SDN, NFV, and 5G protocols into attack graphs. 
We analyze these graphs to generate 119 novel possible exploits that are exclusive 
to 5G networks. They are \textit{possible exploits} in a specific system.
The numerous vulnerabilities arising due to implementation errors are generally system-specific.
We show how these possible exploits can compromise the 5G Authentication and Key Agreement (AKA) protocol. 
We discover five new attack vectors in the 5G-AKA protocol that can be triggered by 5GCN 
vulnerabilities. We demonstrate how various attacks at the network and protocol levels can be 
combined to remotely hack targeted end-user applications. In  a case study, we demonstrate the 
hacking of the WhatsApp account of an end user. We chose WhatsApp as our target application 
because it is the most widely used instant messaging (IM) platform and possesses some of the most 
advanced security features~\cite{sutikno2016whatsapp}. We discovered four security loopholes 
that may be triggered in WhatsApp in the absence of appropriate 5GCN security measures.  We show 
how our framework can scale to larger infrastructures through the use of machine learning (ML) 
and a constraint satisfaction problem (CSP) formulation. We use ML and CSP formulation at the 
system level to predict possible vulnerability exploits when a new node is added to the attack 
graphs. A new node may be added when a new vulnerability is discovered or when a new vulnerable 
component is introduced in the 5GCN. Utilization of ML at the system level is inspired by the 
SHARKS framework~\cite{saha-sharks}, where ML was used to discover novel possible exploits in an IoT 
system. SHARKS is an acronym for Smart Hacking Approaches for RisK Scanning.  Although SHARKS 
was originally targeted at IoT and cyber-physical systems, it is also applicable to the 5GCN 
architecture.

The new contributions of this article include:

\begin{enumerate}
\item Representation of 113 documented SDN and NFV attack vectors in the form of concise attack 
graphs.  
\item Analysis of attack graphs to obtain 119 novel possible exploits of SDN, NFV, and malicious 
peripheral vulnerabilities in the 5GCN.
\item Analysis of the consequences of network infrastructure threats and their interactions on 
the 5G-AKA protocol, resulting in the discovery of five novel possible attack vectors that are 
triggered by 5GCN vulnerabilities.
\item Combination of threats across the hardware, software, network, and protocol layers to 
compromise end-user applications.
\item Application of ML and CSP models to the attack graphs to make the framework scalable to 
larger infrastructures.
\end{enumerate}

The article is organized as follows. Section \ref{sec:Related_work} provides a summary of the 
work that has been done on 5G security. Section \ref{sec:Background} discusses background 
material. 
%Section \ref{sec:Motivation} provides motivation behind why our contribution may be beneficial 
%to progress in 5G security research. 
Section \ref{sec:Methodology} gives details of our methodology. Section \ref{sec:5GAKA} 
describes the impact of system vulnerabilities on the implementation of the 5G-AKA protocol. 
Section \ref{sec:WhatsApp} describes the application of our approach to exploitation of 
network-level vulnerabilities to compromise end-user applications. Section
~\ref{sec:Discussion} includes a discussion on the applications and limitations of our framework.
Section \ref{sec:Conclusion} concludes the article.

\section{Related Work}
\label{sec:Related_work}

Security and privacy of users are of prime importance in 5G networks. The Third Generation 
Partnership Project (3GPP) has been working continuously to define the security standards of 5G 
communication systems.  Multiple versions of security standards have been published to date. 
Recent surveys and articles list the potential vulnerabilities of various 5G-enabling 
technologies like cloud radio access networks, SDN, NFV, network slicing, cloud computing, and 
multi-edge computing~\cite{ahmad20175g,ahmad2018overview,schneider2015towards}. 

There are many vulnerabilities that exist in the SDN ecosystem~\cite{chica2020security}. 
Multiple implementation vulnerabilities exist in various open-source SDN controllers and 
network operating systems (NOSs) like OpenFlow, POX, and OpenDaylight~\cite{yoon2017flow}. 
Similarly, network slicing and NFV have their own vulnerabilities~\cite{lal2017nfv, 
cunha2019network}. NFV inherits many of its vulnerabilities from traditional virtualization 
technologies. However, prior research does not report on the specific attack vectors that can 
exploit these vulnerabilities in the 5G framework and lacks detailed analyses of the impact of 
these vulnerabilities on the end-user. To the best of our knowledge, no prior work explores 
interactions among vulnerabilities of different technologies, like SDN and NFV, to generate 
complex attack vectors. 

We use attack graphs to analyze 5GCN security. Attack graphs have found extensive use in network 
security, software, and electronic systems.
Various vulnerability assessment tools have been developed to analyze the security of software 
systems and networks using attack graphs. Some of the popular ones are MulVal and 
A2G2V~\cite{ou2005mulval, al2019a2g2v}.  However, 
these tools do not address discovery of unique vulnerability exploits in a software-defined and 
virtualized network. We target this problem in this article. ML-based attack graphs have been 
used previously to analyze the security of IoT and cyber-physical systems~\cite{saha-sharks, 
brown-gravitas}. We use ML on the attack graphs to enable our framework to scale to larger 
networks.

The 5G ecosystem consists of multiple protocols executing at different layers. Many 
vulnerabilities have been detected in various 5G protocols like cellular paging 
protocols,
%~\cite{hussain2019privacy}, 
multiple control layer protocols~\cite{hussain20195greasoner}, and cellular access network 
protocols~\cite{borgaonkar2019new}. The 5G-AKA protocol claims to provide higher security than 
its predecessors because it provides enhanced user identity protection, more sophisticated key
derivation, and an increased influence of the home network in authentication. However, the increased complexity of the 5G-AKA 
protocol leads to new vulnerabilities~\cite{dehnel2018security}. Most of these vulnerabilities 
have been detected using formal verification methods~\cite{basin2018formal, 
cremers2019component}. In this article, we investigate how SDN, NFV, and other infrastructure 
vulnerabilities can facilitate the execution of protocol-level attacks. 

In a case study, we analyze the impact of 5G network-level vulnerabilities on the 
implementation of WhatsApp on a client device. WhatsApp is one of the most widely used instant 
messaging platforms with one of the most secure platforms. Due to its 
high popularity and highly secure platform, we choose to examine its security features through 
the lens of a vulnerable network.  Although WhatsApp is highly secure, it is still vulnerable 
to attacks like media file jacking, non-blocking behavior exploitation, voicemail-based 
verification exploits~\cite{wa-voicemail}, and key hijacking attacks. We demonstrate that the 
execution of these attacks becomes easier when we have a compromised 5G network.

\section{Background}
\label{sec:Background}
We analyze the vulnerabilities of various disruptive technologies like NFV, SDN, and network 
slicing. In this section, we provide an introduction to these concepts. We also introduce some 
of the techniques we use to analyze system security.

\subsection{NFV}
A network comprises various network functions (NFs) like gateways, load balancers, and firewalls.
In traditional networks, these NFs are implemented on proprietary hardware systems. Such
systems are not flexible and incur high maintenance costs because they are vendor-proprietary. 
Moreover, they often remain underutilized. These issues prevent network operators from improving 
their average revenue per user (ARPU). NFV provides a way to increase ARPU by reducing network 
CAPEX and OPEX.

NFV abstracts out lower-level NF details by implementing NFs on virtual machines (VMs). This 
facilitates easier adoption of NFs by various applications. In addition, the virtual network 
functions (VNFs) provide higher flexibility and higher resource utilization.

\begin{figure}[h]
\centerline{\includegraphics[scale=0.45]{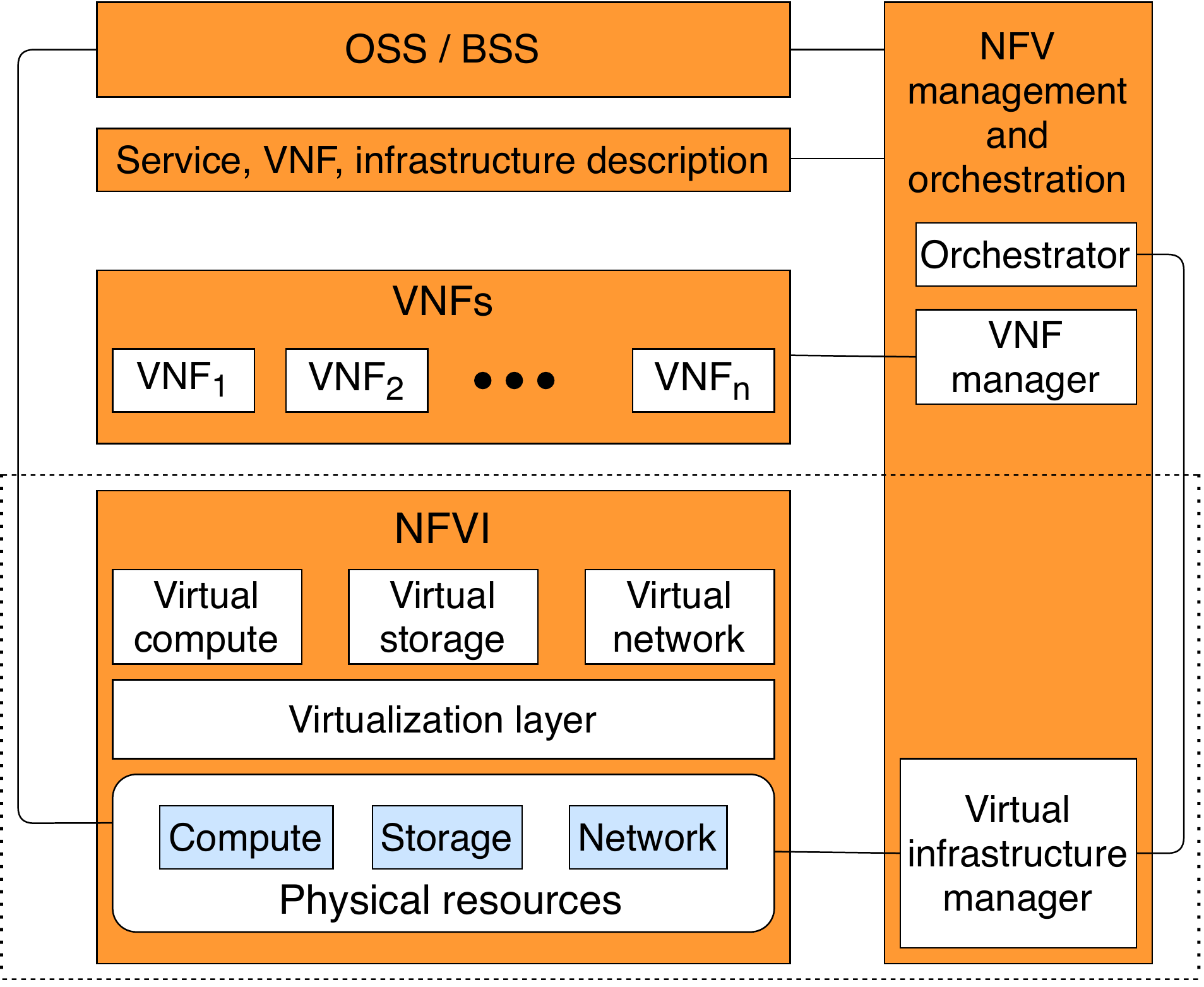}}
\caption{NFV reference architecture provided by the European Telecommunications Standards 
Institute (ETSI)}
\label{fig:nfv-arch}
\end{figure}

The NFV architecture is shown in Fig.~\ref{fig:nfv-arch}. In this figure, every layer interacts 
only with the layers directly above and below it. The rest of the infrastructure is abstracted 
out. For example, the VNFs interact only with OSS/BSS above and virtual resources below. They 
do not need to interact directly with any other layer.

The various components of the NFV architecture are as follows:
\begin{itemize}
    \item \textbf{Operations Support System (OSS):} This is responsible for various network 
management and operations functions like service provisioning and fault tolerance.

    \item \textbf{Network functions virtualization infrastructure (NFVI):} This is a distributed 
system of resources designed to provide a common platform to the VNFs. As shown in 
Fig.~\ref{fig:nfv-arch}, the NFVI can be categorized into three classes: virtual resources, 
virtualization layer, and physical resources.
    
    \item \textbf{NFV orchestrator:} This is part of the NFV management and network 
orchestration (MANO) unit. It plays an important role in instantiating the network.
    
    \item \textbf{VNF manager:} This is responsible for instantiating the VNFs. It manages 
various attributes of the VNFs like their creation, migration, resource allocation, and 
termination.
    
    \item \textbf{Virtual Infrastructure Manager (VIM):} This is responsible for management and 
virtualization of the physical compute, storage, and network resources.
    
\end{itemize}

All the components described above are provided by third-party vendors, unlike pre-5G networks 
where all components are proprietary. This makes these components inherently untrustworthy. Moreover, third-party software systems cannot be protected by hardware-based fingerprinting mechanisms like hardware root-of-trust and physical unclonable functions~\cite{sehwag2016tv}. In 
this article, we study various methods for compromising the virtualization components and the consequences of
doing so.

\subsection{SDN}
Traditionally, network devices have their functionalities hard-coded into the devices. This 
hinders flexibility and innovation in networks. SDNs ameliorate these issues, make 
virtualization of networks easier, and have the potential to increase the ARPU of network 
operators.

The primary objective of software-defined networking is decoupling the control and data planes. 
SDNs have centralized controllers that make forwarding decisions for the switches. The 
controllers have a broad overview of the entire network, hence can make better decisions than 
localized switches.

\begin{figure}[h]
\centerline{\includegraphics[scale=0.57]{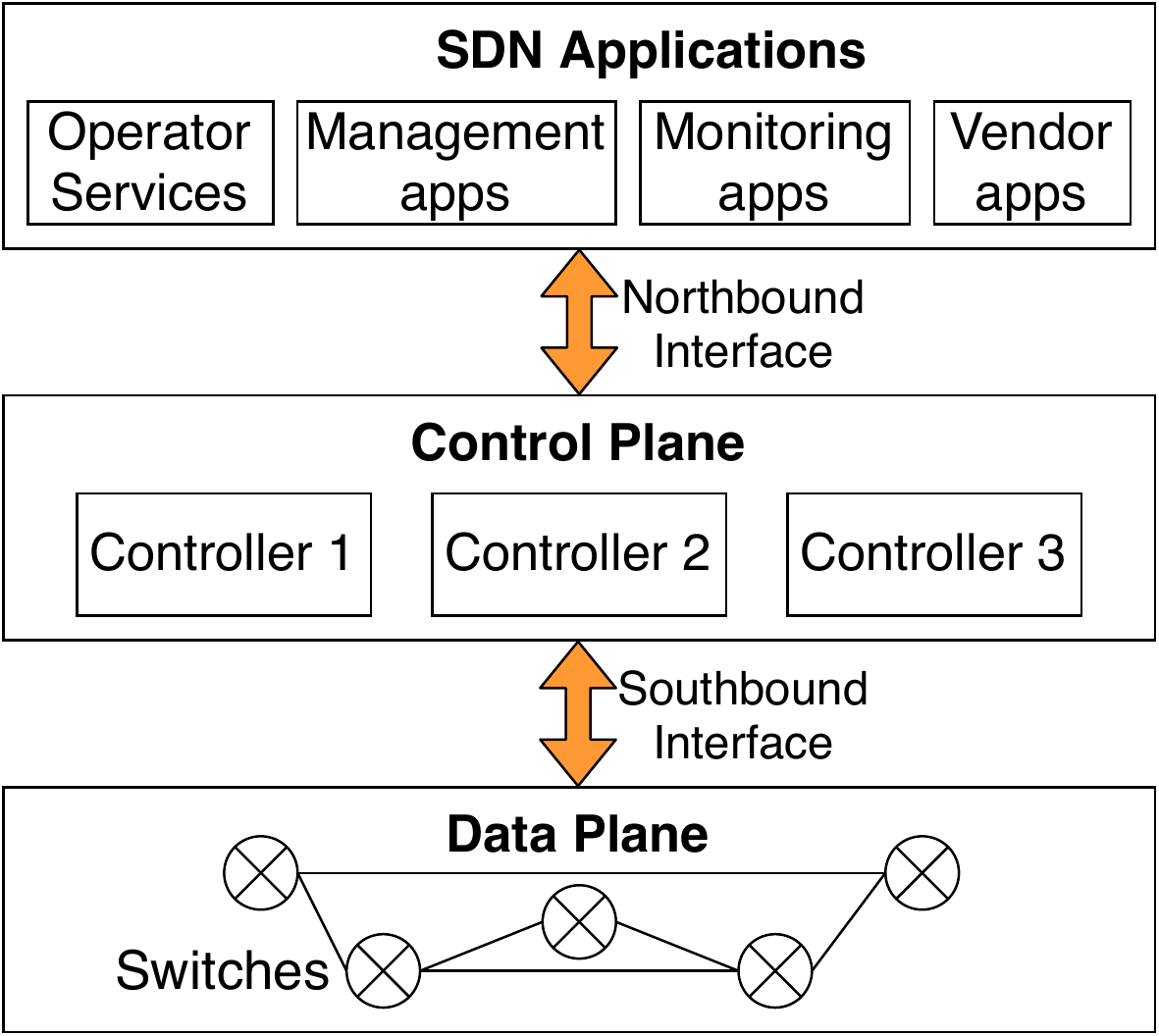}}
\caption{The SDN architecture}
\label{fig:sdn-arch}
\end{figure}

The SDN architecture is shown in Fig.~\ref{fig:sdn-arch}. The logically centralized controllers 
receive application requirements through the northbound interface. They are responsible for 
translating the application requirements into efficient flow rules. These rules are relayed 
to the data plane devices via the southbound interface. The data plane mainly consists of 
forwarding devices like routers and switches. The data plane devices communicate periodically 
with the controllers, updating them with the current situation in the data plane. This gives the 
controllers a global view of the network, thus enabling them to make efficient forwarding 
decisions. 

Various components of the SDN architecture, namely the control plane, data plane, and northbound 
and southbound interfaces, are prone to vulnerabilities. We analyze the consequences of 
these vulnerabilities on a 5G-enabled system.

\subsection{Network Slicing for 5G Networks with SDN/NFV}
\begin{figure}[h]
\centerline{\includegraphics[scale=0.5]{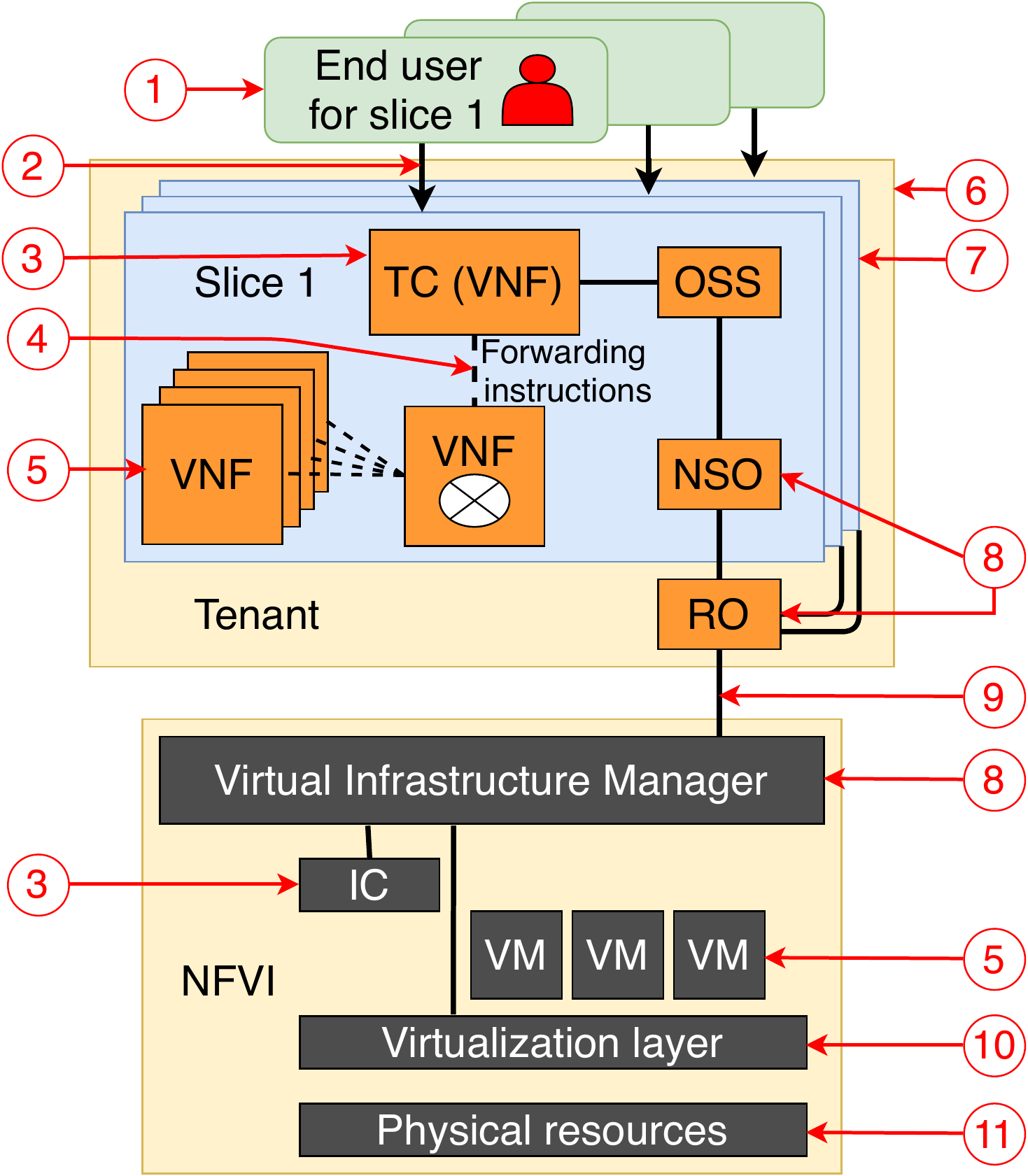}}
\caption{Network slicing for 5G networks with SDN/NFV and its attack surface}
\label{fig:threat-model}
\end{figure}

Network slicing is a method of sharing virtual network resources among multiple verticals. A 
network slice refers to an independent, end-to-end network composed of virtual resources. 
Network slicing enables the network operators to meet their ambitious goals, like scalability 
and low latency, by providing better network isolation and increased statistical multiplexing. 
The network slicing architecture for the 5GCN is depicted in Fig.~\ref{fig:threat-model}.

There are two kinds of resources available for sharing: NFs and the physical infrastructure 
\cite{ordonez2017network}. The NFs are provided to the operators by the tenants and the 
infrastructure by the infrastructure provider (InP). Virtualization and SDN are 
utilized at both the tenant and InP levels. The SDN controller at the tenant and InP levels are 
referred to as the tenant controller (TC) and the infrastructure controller (IC), respectively. 
A simplified example of the implementation of network slicing is depicted in 
Fig.~\ref{fig:threat-model}. Every network slice has a network services orchestrator (NSO) that 
communicates with the resource orchestrator (RO) of the tenant. A tenant provides multiple 
slices to the operators. In the simplified example depicted in Fig.~\ref{fig:threat-model}, 
the tenant is dependent on a single InP for its resources. In reality, the tenant may be 
dependent on multiple InPs.

The network slicing architecture demonstrates how SDN and NFV are used together in the 5GCN. 
We use this framework while analyzing 5G system security. 

\subsection{Regular Expression}
Regular expression is a concise representation of a set of strings. We use regular expressions 
to represent an attack vector.  
The set of all permissible characters in a regular expression is called its alphabet, denoted by 
$\Sigma$. The operations in regular expressions that we use in this article are described in 
Table \ref{table:regex}. 

\begin {table}[h]
\caption {The basic operations in our regular expressions}
\begin{tabular}{m{1.7cm} m{2.5cm} c}
\hline
& & $A=\{a,b\}; B=\{c,d\}$ \\ \hline
\textbf{Operation} & \textbf{Definition} & \textbf{Example}\\ \hline
Set Union (+) & Set union of two regular expressions & $A+B=\{a,b,c,d\}$\\ \hline
Concatenation (.) & Concatenation of strings of two regular expressions & $A.B=\{ac,ad,bc,bd\}$\\ \hline
\end{tabular}
\label{table:regex}		
\end {table}

Regular expressions are generally used to denote system-level operations that are
incomprehensible to humans. In this article, we define the characters of the regular expression 
at a higher granularity for the sake of generality. The alphabet ($\Sigma$) of our regular 
expressions comprises human-understandable system-level operations. For example, 
$\Sigma=$\{'Install malicious switch,' 'Insert malware in hypervisor,' ...\}. This is done to 
ensure that application of our approach is independent of the application, OS or the compiler 
employed by the 5GCN.

\section{Methodology}
\label{sec:Methodology}
This section describes our methodology and its impact. We analyze the security of the
software-defined and virtualized 5GCN using ML and CSP formulation. Section 
\ref{sec:Meth_threat_model} describes our threat model.  Section \ref{sec:Meth_AV_rep} 
describes the method of representing attack vectors with attack graphs.  Section 
\ref{sec:Meth_5GCN_VA} gives details of analyzing 5GCN security with attack graphs.  Section 
\ref{sec:Meth_ML_DAG} describes the methods for exploiting ML and CSP formulation to improve 
the scalability of the proposed methodology.

\subsection{Attack Surface}
\label{sec:Meth_threat_model}

An attack surface of a system refers to the set of various entry points that can be exploited.
The various components that compose the attack surface of the 5GCN 
are depicted in Fig.~\ref{fig:threat-model}. They are as follows:

\begin{enumerate}
    \item User applications
    \item Northbound interface of SDN controller
    \item SDN controller
    \item Control channel of SDN
    \item VNFs
    \item Tenant
    \item Network slice
    \item NFV MANO unit
    \item Management network between tenant and InP
    \item Hypervisor
    \item InP peripheral attacks; Attacks on physical infrastructure
\end{enumerate}

The attack vectors for exploiting vulnerabilities of these components are discussed in detail 
in the subsequent sections.

\subsection{Attack Vector Representation}
\label{sec:Meth_AV_rep}
We use regular expressions and attack graphs to represent various attacks on the 5G system. We use regular expressions because they allow us to represent the sequence of exploits in an exploit chain.
We use attack graphs because they enable efficient modeling of the interactions between different 
threats. In this section, we describe the process of constructing the attack graphs from 
various attacks.  First, every attack is decomposed into a sequence of system-level operations. 
We represent this sequence using a regular expression. Then, we convert this regular expression 
into an attack graph.  For example, let us consider an attack in which a target switch is 
disconnected from its SDN controller by poisoning the Address Resolution Protocol (ARP). This 
attack can be executed by the following sequence of system-level operations:

\begin{enumerate}
    \item Install a malicious VM in the system.
    \item Launch an ARP poisoning attack to alter the MAC address of the controller on the 
target switch.
    \item  In the target switch memory, replace the MAC address of the original controller with 
that of the malicious VM.
    \item The target switch is now disconnected from the controller.
    \item Send malicious flow rules to the switch from the malicious VM. This disrupts network 
functionalities.
\end{enumerate}

\begin{figure}[h]
\centerline{\includegraphics[scale=0.55]{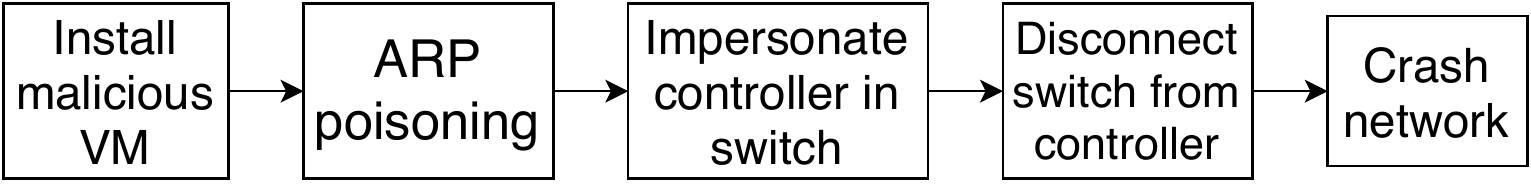}}
\caption{Turning a regular expression into an execution graph}
\label{fig:regex-exa}
\end{figure}

Let $ch$ denote a character from the alphabet $\Sigma$ of our regular expressions. Then, the 
regular expression of the attack vector described above can be represented as: 
$ch_i$(Install malicious VM). $ch_j$(ARP poisoning).  $ch_k$(Impersonate controller in switch). 
$ch_l$(Disconnect switch from controller). $ch_m$(Crash network).  This regular expression can 
be converted into an execution graph, as shown in Fig.~\ref{fig:regex-exa}.

We combine the execution graphs of multiple attacks to obtain the aggregated attack graphs.

\subsection{5GCN Vulnerability Analysis}
\label{sec:Meth_5GCN_VA}
In this section, we describe the vulnerabilities of NFV, SDN, and peripheral devices, and the threats 
that arise from them.  For each of these domains, we use regular expressions and attack graphs to 
conduct a complete security analysis of the system.

\subsubsection{SDN Vulnerability Analysis}
SDN is one of the most disruptive technologies that is deployed in 5G systems. SDN implementation contains multiple vulnerabilities and is prone to exploits of varying complexity, including topology poisoning attacks, controller hijacking attacks, man-in-the-middle (MiTM) attacks, and 
denial-of-service (DoS) attacks, to name a few \cite{ahmad2015security, kreutz2013towards, 
scott2015survey, chica2020security}. Moreover, popular open-source NOSs for the SDN controller, 
namely OpenFlow, OpenDaylight, and POX, have been shown to be vulnerable to multiple attacks 
\cite{yoon2017flow}.

\begin{figure*}[h]
\centerline{\includegraphics[scale=0.45]{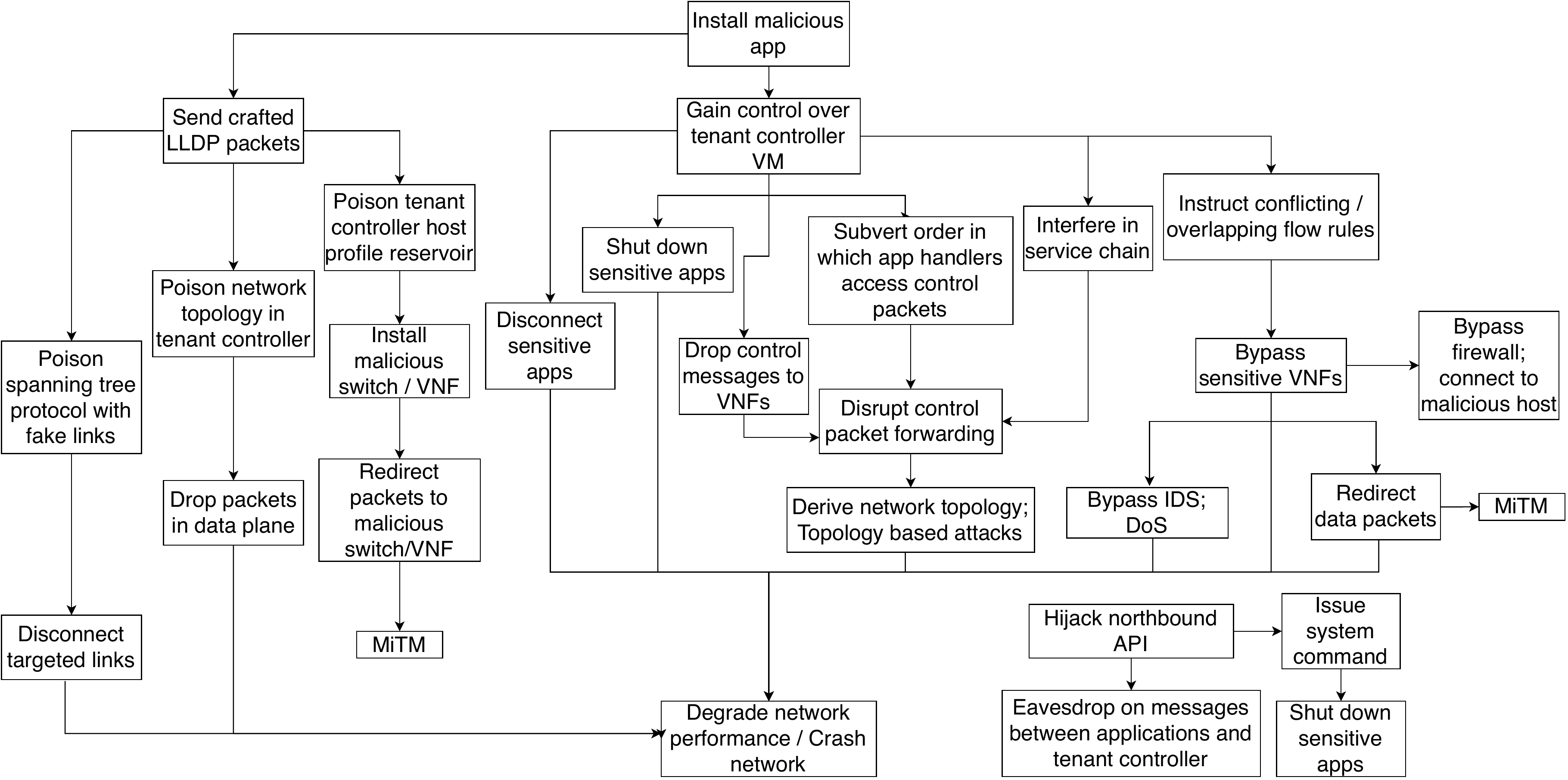}}
\caption{Aggregated attack graph of SDN control plane vulnerabilities}
\label{fig:sdn-cp}
\end{figure*}

\begin{figure*}[h]
\centerline{\includegraphics[scale=0.45]{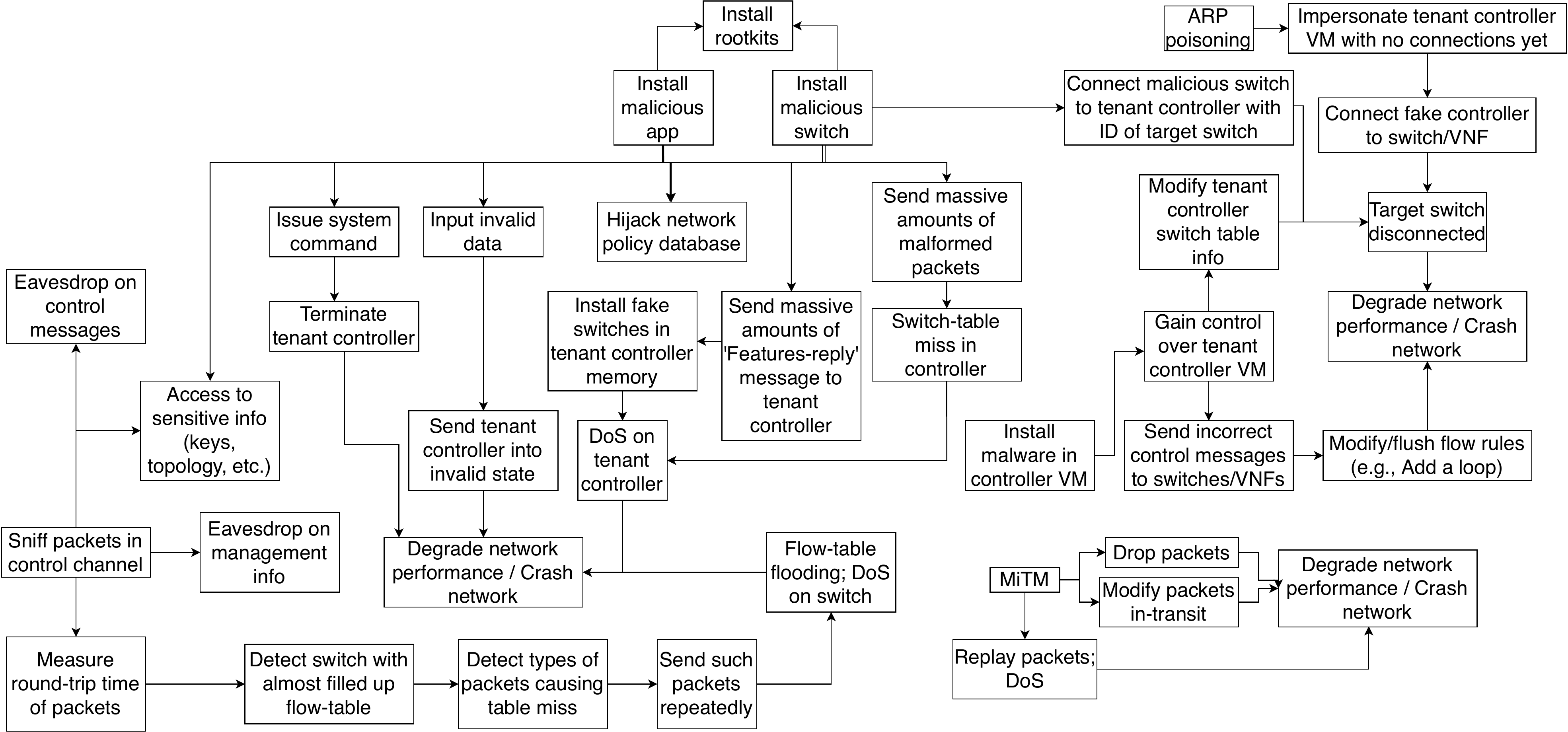}}
\caption{Aggregated attack graph of SDN data plane vulnerabilities}
\label{fig:sdn-dp}
\end{figure*}

SDN vulnerabilities can be broadly divided into two categories: control plane and data plane 
attacks.  Control plane attacks involve compromising the NOS or the control channel that is 
used to send the control messages to the data plane devices \cite{cao2019crosspath}. The 
communication in the control channel is generally unencrypted to enhance performance. This is a 
potential security loophole. An adversary with access to the control channel can possibly 
eavesdrop on the control messages to infer the network topology. Knowledge of the network 
topology can lead to a variety of attacks~\cite{alharbi2015security, marin2019depth}. Moreover, 
an adversary can compromise the integrity of the control messages without being detected. This 
can cause malicious network reconfiguration and DoS attacks. The data plane is also vulnerable 
to various attacks. The data plane attacks generally target individual switches and forwarding 
devices. We represent all SDN attack vectors as regular expressions and then convert them into 
attack graphs. The regular expressions of various SDN attacks are shown in 
Table~\ref{tab:regex-sdn}. These attack vectors in the SDN control plane (SDN-CP) and the SDN 
data plane (SDN-DP) are then concisely represented as the attack graphs shown in 
Fig.~\ref{fig:sdn-cp} and Fig.~\ref{fig:sdn-dp}, respectively. In an attack graph, every path 
from a head node to a tail node is a unique attack vector. The graph in Fig.~\ref{fig:sdn-cp} 
has 14 unique SDN-CP attack vectors and the graph in Fig.~\ref{fig:sdn-dp} has 25 unique SDN-DP 
attack vectors.

\begin{table*}[]
\caption{Regular Expressions for SDN attacks~\cite{chica2020security, ahmad2015security, kreutz2013towards, scott2015survey, cao2019crosspath, alharbi2015security, marin2019depth}}
\begin{tabular}{lll}
%& & \textbf{Regular Expression} \\
\hline
%\multicolumn{2}{l}{\textbf{Application layer}} & \textbf{Regular Expression}\\ \hline \hline
\textbf{Application layer}& \textbf{Entry point of attack} & \textbf{Regular expression}\\ \hline \hline
\begin{tabular}[c]{@{}l@{}}Abuse of privileges \\ and authority\end{tabular} & Malicious third party 
apps & \begin{tabular}[c]{@{}l@{}}$ch_i$(install malicious app). $ch_n$(gain control over tenant controller VM). \\ \{$ch_j$(disconnect sensitive apps) + $ch_k$(shutdown sensitive apps)\}. \\ \{$ch_l$(crash network) + $ch_m$(degrade network performance)\}\end{tabular}\\ \hline

Service disruption & Malware & \begin{tabular}[c]{@{}l@{}}$ch_i$(install malicious app). $ch_j$(gain control over tenant controller VM). \\ {[}\{$ch_1$(drop control messages to VNFs) + $ch_2$(subvert order in which app \\ handlers access control packets) + $ch_3$(interfere in service chain)\}. \\ $ch_k$(disrupt control packet forwarding) + $ch_4$(eavesdrop on control messages). \\ $ch_l$(derive topology of network). $ch_m$(execute topology based attacks){]}.\\ \{$ch_o$(crash network) + $ch_p$(degrade network performance)\}\end{tabular} \\ \hline
Application shutdown & Vulnerable northbound API & \begin{tabular}[c]{@{}l@{}}$ch_i$(exploit vulnerability in northbound API). \{$ch_j$(issue system command). \\$ch_k$(terminate victim app) + $ch_l$(eavesdrop on messages between controller and app)\}\end{tabular} \\ \hline
\multicolumn{3}{l}{\textbf{Control layer}} \\ \hline \hline

\begin{tabular}[c]{@{}l@{}}Dynamic flow rule \\ tunneling\end{tabular} & \begin{tabular}[c]{@{}l@{}}Malware \& vulnerable \\ switches\end{tabular} & \begin{tabular}[c]{@{}l@{}}$ch_i$(install malicious app). $ch_j$(instruct conflicting/overlapping flow rules). \\ $ch_k$(bypass  sensitive VNFs like firewall/Intrusion Detection Systems (IDS)).\\ \{$ch_l$(degrade performance) + $ch_m$(crash network) + $ch_n$(DoS attack)\}\end{tabular} \\ \hline
\begin{tabular}[c]{@{}l@{}}Controller poisoning \\ (Poisoned network view)\end{tabular} &
\begin{tabular}[c]{@{}l@{}}Malware \& vulnerable \\ network services and \\
protocols\end{tabular}     & \begin{tabular}[c]{@{}l@{}}$ch_i$(gain access to controller VM).\\
\{$ch_j$(send crafted LLDP packets). $ch_k$(poison network topology in controller \\ by adding
fake connections). $ch_l$(drop packets in data plane). $ch_m$(degrade \\ performance) +
$ch_n$(poison controller host profile reservoir). $ch_o$(install malicious\\  VM). $ch_p$(redirect data packets to malicious VM). $ch_q$(MiTM attack in data plane)\}\end{tabular} \\ \hline
NOS misuse & Vulnerable controller & \begin{tabular}[c]{@{}l@{}}\{$ch_i$(malicious apps running at Application layer) + $ch_j$(rogue switch VM)\}. \\ (multiple attacks that are denoted in cells below)\end{tabular} \\ \hline
& & \begin{tabular}[c]{@{}l@{}}$ch_k$(execute system commands). $ch_l$(terminate controller). \\ $ch_m$(degrade network performance)\end{tabular}                  \\ \hline
& & $ch_k$(access sensitive network information). $ch_l$(execute deviant actions)                                
\\ \hline
& & \begin{tabular}[c]{@{}l@{}}$ch_k$(modify flow rules). \{$ch_l$(eavesdrop on data plane packets) + $ch_m$(redirect \\ data packets)\}. \{$ch_n$(degrade network performance) + $ch_o$(MiTM attack in \\ data plane) + $ch_p$(bypass security functions like firewalls/IDS)\}\end{tabular}              \\ \hline
& & $ch_k$(install rootkits)               \\ \hline
& & $ch_k$(hijack network policy database) \\ \hline
& & \begin{tabular}[c]{@{}l@{}}$ch_k$(input invalid input data). $ch_l$(send controller in an invalid state). \\ $ch_m$(degrade network performance)\end{tabular}          \\ \hline
Packet-in flooding & \begin{tabular}[c]{@{}l@{}}Faulty controller or\\ compromised switch
VMs\end{tabular} & \begin{tabular}[c]{@{}l@{}}\{$ch_i$(malicious app) +
$ch_j$(malicious switch VM)\}. $ch_k$(send massive amounts of \\ malformed packets). $ch_l$(switch-table misses of switch VM). $ch_m$(massive\\  amount of packet-in messages sent to controller VM). $ch_o$(DoS attack on\\  controller). $ch_p$(degrade network performace)\end{tabular}           \\ \hline
Switch table flooding & \begin{tabular}[c]{@{}l@{}}Faulty controller or \\ compromised switch VMs\end{tabular} & \begin{tabular}[c]{@{}l@{}}\{$ch_i$(malicious app) + $ch_j$(malicious switch VM)\}. $ch_k$(send massive amount of \\ 'features-reply' messages to controller). $ch_l$(fill controller switch table with \\ fake switches).  $ch_m$(DoS attack on controller). $ch_n$(degrade network performance)\end{tabular}                  \\ \hline
\begin{tabular}[c]{@{}l@{}}Legitimate switch\\ id hijacking\end{tabular} &
& \begin{tabular}[c]{@{}l@{}}$ch_i$(malicious switch VM installed). $ch_j$(connect malicious VM to controller using \\DPID of target VM). $ch_k$(legitimate VM gets disconnected). \{$ch_l$(network crash) + \\ $ch_m$(degrade network performance)\}\end{tabular}                \\ \hline
Spanning tree poisoning & & \begin{tabular}[c]{@{}l@{}}$ch_i$(send crafted LLDP packets to controller VM). $ch_j$(poison
spanning tree \\ protocol with targetted fake links). $ch_k$(disconnect targetted links). \\ \{$ch_l$(network crash) + $ch_m$(degrade network performance)\}\end{tabular}                \\ \hline
%Malware on NOS & & \begin{tabular}[c]{@{}l@{}}$ch_i$(install malware in controller VM). $ch_j$(modify controller switch table info). \\ $ch_k$(disconnect a switch VM). \{$ch_l$(network crash) + $ch_m$(degrade network \\ performance)\}\end{tabular}     \\ \hline
\multicolumn{3}{l}{\textbf{Control Channel (CC)}}                                     \\ \hline \hline
Passive MiTM & Unencrypted messages & \begin{tabular}[c]{@{}l@{}}$ch_i$(absence of crypto in CC). $ch_j$(sniff packets on CC). \{$ch_k$(eavesdrop on \\ control messages) + $ch_l$(eavesdrop on topology information) + \\ $ch_m$(eavesdrop on management 
info)\}\end{tabular}                       \\ \hline
Active MiTM & \begin{tabular}[c]{@{}l@{}}Compromised southound \\ interface or vulnerable \\ data links\end{tabular} & \begin{tabular}[c]{@{}l@{}}$ch_i$(absence of crypto in CC). $ch_j$(ARP poisoning). $ch_k$(insert intruder host \\ between controller and data plane)\end{tabular}    \\ \hline
\multicolumn{3}{l}{\textbf{Infrastructure layer}}                                    \\ \hline \hline
\begin{tabular}[c]{@{}l@{}}DoS leveraging \\ ARP poisoning\end{tabular} & & \begin{tabular}[c]{@{}l@{}}$ch_i$(ARP poisoning). $ch_j$(impersonate controller VM). $ch_k$(connect fake \\ controller to target switch). $ch_l$(disconnect target switch VM from network). \\ $ch_m$(degrade network performance)\end{tabular}        \\ \hline
\begin{tabular}[c]{@{}l@{}}Flow-rule flushing/\\ modification\end{tabular} & & \begin{tabular}[c]{@{}l@{}}\{$ch_i$(malicious app) + $ch_j$(install malware on controller VM) + $ch_o$(gain access to \\ controller VM)\}. $ch_k$(send incorrect control messages to switches). \\ \{$ch_l$(modify switch flow rules) + $ch_m$(flush switch flow rules)\}. $ch_n$(degrade \\ network performance)\end{tabular}                  \\ \hline
Flow-rule flooding & Side-channel attack (SCA) &
\begin{tabular}[c]{@{}l@{}}$ch_i$(record round-trip time of packets; SCA). $ch_j$(detect \\ VM that has an almost full switch-table). $ch_k$(detect types of packets causing \\ table misses). $ch_l$(send such packets repeatedly). $ch_m$(flood switch table of VM). \\ $ch_n$(degrade network performance)\end{tabular}                  \\ \hline
\end{tabular}
\label{tab:regex-sdn}
\end{table*}

\subsubsection{NFV Vulnerability Analysis}
NFV provides a dynamic and loosely-coupled infrastructure that caters to a large diversity of 
user requirements. However, NFV inherits multiple implementation vulnerabilities and exploits 
thereof.  Prior to NFV, when a proprietary function was introduced in the network, there 
existed an established trust between the developer and the operator. This trust is absent in an 
NFV-enabled network architecture because third-party VNFs are usually susceptible to a variety 
of threats \cite{nfv2014etsi}:

\begin{enumerate}
    \item Generic networking threats.
    \item Generic virtualization threats.
    \item Emerging threats due to a combination of networking and virtualization. 
\end{enumerate}

Due to multi-tenancy and Infrastructure-as-a-Service paradigms of virtualization, access to the 
core network is easier than before. This makes the 5GCN vulnerable to different kinds of 
attackers, some of whom may be end customers of retail networks, retail network operators, 
wholesale network operators, hypervisor operators, infrastructure sharers and operators, and 
facility managers. Hence, security monitoring should be an integral part of the 5GCN ecosystem.

\begin{figure*}[h]
\centerline{\includegraphics[scale=0.45]{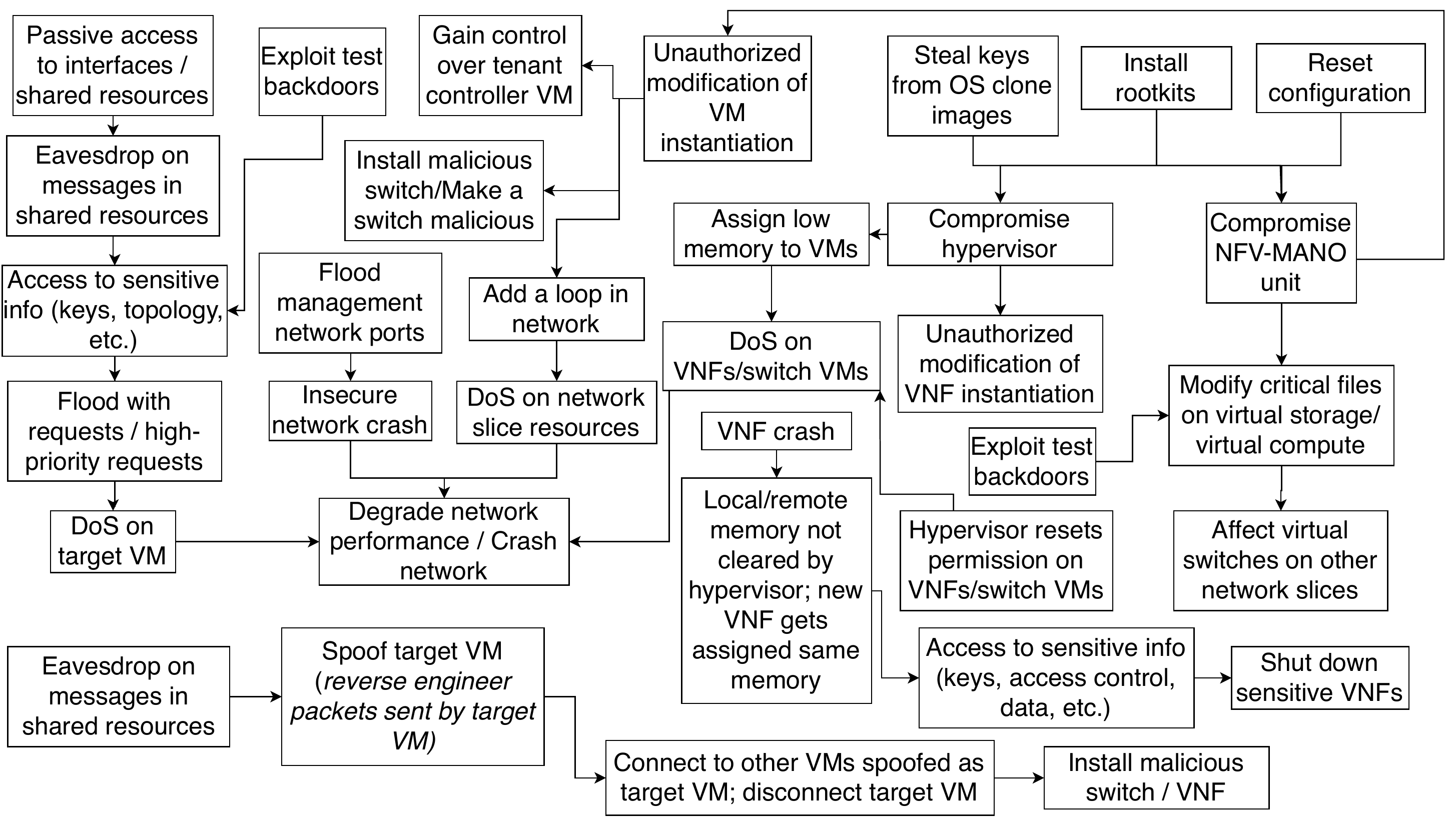}}
\caption{Aggregated attack graph of NFV vulnerabilities}
\label{fig:nfv-threat}
\end{figure*}

The regular expressions of the NFV threats and vulnerabilities are described in 
Table~\ref{tab:regex-nfv} and are concisely represented in the attack graph shown in 
Fig.~\ref{fig:nfv-threat}. The attack graph has 25 unique NFV attack vectors. These attack 
vectors have been constructed from the ETSI NFV security problem statement~\cite{nfv2014etsi}.

\begin{table*}[]
    \centering
    \caption{Regular expressions of NFV threat vectors mentioned in the standards document ETSI\_GS\_NFV-SEC\_001\_v1.1.1 \cite{nfv2014etsi}}
    \begin{tabular}{p{3cm} p{14cm}}
    \hline
     \textbf{Topology-based attacks} & \textbf{Regular expression} \\
     \hline \hline
     %\multicolumn{2}{l}{\textbf{Topology-based attacks}} \\ \hline \hline
     Adding unauthorized connection in VNF & $ch_i$(modify VNF instantiation). $ch_j$(add
unauthorized connection in VNF). $ch_k$(exploit weak crypto implementations).
\{$ch_l$(eavesdrop on packets) + $ch_o$(add a loop in network). $ch_p$(orchestrator creates new
instances of VMs (to handle excess load). $ch_q$(DoS on NFV infrastructure)\} \\ \hline
     Modifying firewall/IDS instantiation & $ch_i$(modify VNF instantiation). {$ch_j$(modify rules
in firewall virtual storage). $ch_k$(connect to malicious website) + $ch_l$(modify IDS rules).
\{$ch_m$(flood network with incoming malicious traffic). $ch_p$(orchestrator creates new
instances of VMs (to handle excess load)). $ch_q$(DoS on NFV infrastructure) + $ch_m$(flood
network with incoming DNS queries). $ch_p$(orchestrator creates new virtual DNS).
$ch_q$(amplified DNS query request - DoS attack on victim)\}} \\ \hline
     Passive MiTM & \{$ch_i$(modify VNF instantiation). $ch_m$(add a link to malicious VM) + $ch_j$(physical access of interfaces)\}. $ch_k$(eavesdrop on the messages being sent). $ch_l$(infer topology of network) \\ \hline
     Active MiTM & $ch_o$(physical access to interface). $ch_i$(exploit weak crypto implementations). $ch_j$(modify packets in-transit). $ch_k$(crash the system) + $ch_l$(replay packets). $ch_m$(orchestrator creates new instances of VMs (to handle excess load)). $ch_n$(DoS on NFV infrastructure) \\ \hline
     \multicolumn{2}{l}{\textbf{Exploiting Lights out Management (LOM)}} \\ \hline \hline
     DoS attack on management network & $ch_i$(identify network port(s) having access to LOM). $ch_j$(flood the port with requests). $ch_k$(DoS attack on LOM port(s)) \\ \hline
     Exploiting LOM network & \{$ch_i$(SQL injection attack on virtual storage) + $ch_j$(SCA by
physical access) + $ch_k$(cache poisoning attack) + $ch_l$(download unwhitelisted software).
$ch_m$(gain control over hypervisor) + $ch_n$(dynamic memory overflow of hypervisor).
$ch_o$(overwrite frame pointer of hypervisor). $ch_p$(code injection in hypervisor)\}.
$ch_q$(read secret LOM credentials).\{$ch_r$(crash the system) + $ch_s$(modify critical files
on virtual storage) + $ch_t$(modify critical code on virtual compute)\} \\ \hline
     Exploiting hypervisor dependency on VNF (1) & $ch_i$(network fails). $ch_j$(hypervisor starts to
boot). $ch_k$(hypervisor requests network configuration from VM running on top of it).
\{$ch_l$(VM crashed; depends on hypervisor2 which has also crashed) + $ch_m$(crashed VM depends
on hypervisor)\}. $ch_n$(VM fails to boot). $ch_o$(DoS on hypervisor) \\ \hline
     Exploiting hypervisor dependency on VNF (2) & $ch_i$(network failure). $ch_j$(virtual forwarding function1 (VFF1) starts to boot). $ch_k$(requests access to VFF2). $ch_l$(VFF2 is crashed). $ch_m$(VFF2 requests access to VFF1). $ch_n$(deadlock arises, DoS on VFF1). $ch_o$(DoS on VFF2) \\ \hline
     Exploiting insecure boot & $ch_i$(absence of secured boot authentication). {$ch_j$(steal secret keys) + $ch_k$(rootkit injection) + $ch_p$(reset configuration)}. {$ch_l$(hypervisor compromise) + $ch_m$(VM compromise) + $ch_n$(orchestrator compromise) + $ch_o$(VM manager compromise)} \\ \hline
     \multicolumn{2}{l}{\textbf{Insecure Crash}} \\ \hline \hline
     Compromising sensitive data & $ch_i$(VM/VNF crashes). \{$ch_j$(local memory not cleared by hypervisor) + $ch_o$(remote memory not cleared by hypervisor)\}. $ch_k$(new VM gets assigned same memory addresses as crashed VM). $ch_l$(new VM gets access to sensitive data) + $ch_m$(new VM gets access to keys). $ch_n$(new VM implements privilege escalation) \\ \hline
     Exploiting absence of safety measures & $ch_i$(application within VM crashes but VM is still functional). $ch_j$(hypervisor resets/changes existing authorizations). $ch_k$(VM is restricted from performing required functions). $ch_l$(network crashes) \\ \hline
     Privilege escalation & $ch_i$($VM_i$ crashes). $ch_j$(memory and authorizations are not cleared by hypervisor). $ch_k$($VM_j$ gets assigned the same memory location as crashed instance of $VM_i$). $ch_l$($VM_j$ gets same privileges as $VM_i$) \\ \hline
     Authentication, Authorization, Accounting (AAA) attacks (1) &  $ch_i$(weak authentication on NFVI manager). $ch_j$(access to hypervisor). $ch_k$(access to physical storage, compute and network). \{$ch_l$(get secret keys) + $ch_m$(MiTM attacks) + $ch_n$(replay attacks) + $ch_o$(eavesdrop on communication) + $ch_p$(modify packets in-transit) + $ch_q$(assign low memory to VMs). $ch_r$(DoS on VMs) + $ch_s$(give unauthorized privileges to malicious actors) + $ch_t$(download unwhitelisted malware) + $ch_u$(add unauthorized connections)\} \\ \hline
     AAA attacks (2) & $ch_i$(weak auth. of hypervisor) + $ch_j$(weak auth. of orchestrator) +
$ch_k$(weak auth. of VM) + $ch_l$(weak auth. of VNF managers)\\ \hline
     Exploiting backdoors meant for testing & $ch_i$(virtualized switch in
promiscuous mode). \{$ch_j$(eavesdrop on VNF traffic with test process) + $ch_k$( adversary
sends malicious traffic through test backdoor) + $ch_l$(shared memory access to test
process)\}.\{$ch_m$(eavesdrop on sensitive data/secret keys) + $ch_n$(modify sensitive data of
VNF in test/monitoring mode)\} \\ \hline
     Flooding attacks & $ch_i$(physical access to shared network resources). \{$ch_j$(flood shared network with requests) + $ch_k$(flood shared network with high-priority messages)\}. $ch_l$(DoS attack on target VM) \\ \hline
     Eavesdropping on shared resources & \{$ch_g$(virtual sharing of same network slice) +
$ch_h$(physical sharing of same network component)\}. $ch_i$(eavesdrop on shared
resources). $ch_i$(absence of crypto on control plane). $ch_j$(reverse engineer the packets sent
by target VM). $ch_k$(spoof target VM). \{$ch_m$(send modified packets with target VM id) +
$ch_n$(request access to other VMs with target VM’s id). $ch_o$(target VM is disconnected from
these VMs in the virtual network). $ch_p$(replay packets received by target VM)\}.\{$ch_q$(crash
the system)+$ch_r$(launch DoS attack on another VM through target VM)\} \\ \hline
     SCA/Cache poisoning & {$ch_i$(SCA analysis) + $ch_j$(cache poisoning)}. $ch_k$(extract crypto keys) \\ \hline
     \multicolumn{2}{l}{\textbf{Resources of virtual infrastructure}} \\ \hline \hline
     Local storage attacks & $ch_i$(install malware at hypervisor level). $ch_j$(force hypervisor to fill up local storage with logs). $ch_k$(local storage insufficient for VMs). \{$ch_l$(degrade network performance) + $ch_m$(DoS attack on VMs)\} \\ \hline
     Remote attacks & $ch_i$(install malware at hypervisor level). \{$ch_j$(force hypervisor to fill up remote storage with logs). $ch_k$(remote storage insufficient for VMs) + $ch_n$(remote control channel degradation)\}. \{$ch_l$(degrade network performance) + $ch_m$(DoS attack on VMs)\} \\ \hline
     Memory pressure attacks & $ch_i$(install malware at hypervisor level). $ch_j$(consume kernel memory). \{$ch_l$(degrade network performance) + $ch_m$(crash the system)\} \\ \hline
     CPU attacks & $ch_i$(install malware at hypervisor level). $ch_j$(cause scheduler unfairness). \{$ch_l$(degrade network performance) + $ch_m$(crash the system)\} \\ \hline
     OS resource exhaustion & $ch_i$(install malware at hypervisor level). \{$ch_j$(consume file handles) + $ch_n$(consume event channels)\}. $ch_o$(insufficient resources for OS). \{$ch_l$(degrade network performance) + $ch_m$(crash the system)\} \\ \hline
    \end{tabular}
    \label{tab:regex-nfv}
\end{table*}

\subsubsection{Malicious Peripherals}
\begin{figure*}[h]
\centerline{\includegraphics[scale=0.45]{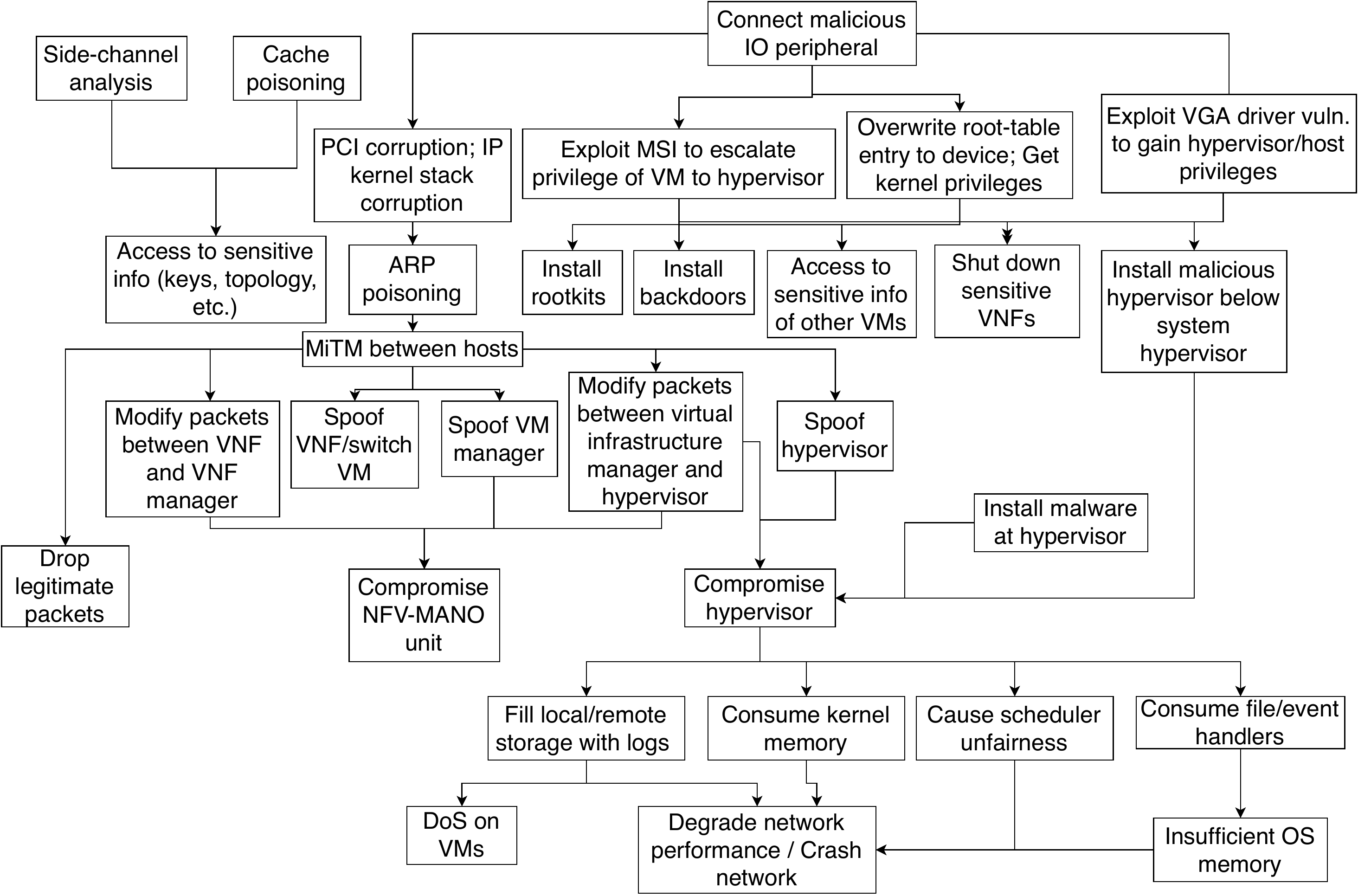}}
\caption{Aggregated attack graph of malicious peripheral based attacks}
\label{fig:mal-per}
\end{figure*}

The 5GCN is vulnerable to malicious peripheral devices that can potentially compromise the 
virtualization infrastructure. Input/Output (IO) attacks involve malicious peripherals that 
make root-level read and write accesses to the DRAM or to the memory embedded in other 
peripherals. Various attacks involve corrupting the Peripheral Component Interconnect (PCI) 
to install rootkits \cite{morgan2018iommu}, exploiting Message Signal Interrupts (MSI) and VGA 
driver vulnerabilities for privilege escalation on hypervisors, and overwriting root-table 
entries to gain kernel privileges. A concise representation of these attacks 
is shown in the attack graph in Fig.~\ref{fig:mal-per}. This attack graph consists of 49 attack 
vectors introduced by malicious peripherals and attacks on physical infrastructure of InP.

The vulnerabilities mentioned in this section require physical access to the infrastructure. 
Hence, they are less likely to be exploited than NFV and SDN vulnerabilities. However, such 
attacks are quite common and their impact is often catastrophic. Thus, it is necessary to 
take precautions against such attacks while designing the system.

\subsubsection{Graph Analysis}
\label{sec:graph-analysis}
The attack vectors in the graphs are constructed from SDN vulnerabilities pointed out in
existing literature~\cite{chica2020security, cao2019crosspath, marin2019depth}, 
NFV vulnerabilities~\cite{nfv2014etsi,lal2017nfv}, and IO vulnerabilities~\cite{morgan2018iommu, 
markettos2019thunderclap, li2019protecting}. We find that there are 113 attack vectors in all
in the four aggregated attack graphs. This is summarized in Table~\ref{tab:graph-attacks}.

\begin{table}[]
    \centering
    \caption{Summary of attacks in the graphs}
    \begin{tabular}{c|c}
     \textbf{Graph} & \textbf{Number of attack vectors} \\
     \hline
     SDN-CP & 14 \\
     SDN-DP & 25 \\
     NFV & 25 \\
     Malicious peripheral & 49 \\
     \hline
     \textbf{Total} & \textbf{113}\\ 
    \end{tabular}
    \label{tab:graph-attacks}
\end{table}

After constructing the attack graphs based on previous literature, we observe that many of the 
unconnected nodes in these graphs can be linked together to generate new possible exploits. In this 
section, we analyze the feasibility of connections among the unconnected nodes. A link or a 
branch is deemed to be feasible if the control/data flow represented by that branch is feasible 
in a real-world system. For example, nodes \textit{'Exploit test backdoors'} and 
\textit{'Access sensitive information'} can be connected because sensitive credentials of a 
resource can be accessed through backdoors. On the other hand, nodes 
\textit{'Compromise hypervisor'} and \textit{'Flood management ports'} cannot be connected 
because there is a lack of a direct causal relationship between the two. 
 
Connecting a pair of nodes leads to a new directed branch in the graph. A new branch is 
interpreted as a novel possible exploit of an existing vulnerability. There are two categories of 
novel possible exploits in this analysis:
\begin{itemize}
    \item \textbf{Intra-graph:} These possible exploits are restricted to one of 
the four domains, namely SDN-CP, SDN-DP, NFV, and malicious peripherals. For example, when we connect 
two nodes in Fig.~\ref{fig:sdn-cp}, we get a novel possible exploit in the SDN-CP.
    \item \textbf{Inter-graph:} These possible exploits involve the combination 
of vulnerabilities of multiple attack graphs. For example, when we connect a node in 
Fig.~\ref{fig:sdn-dp} to a node in Fig.~\ref{fig:nfv-threat}, it leads to a novel possible exploit 
that combines vulnerabilities of the SDN-DP with that of the NFV infrastructure.
\end{itemize}

We demonstrate some of our novel possible exploits in Table~\ref{tab:exa-exploits}. We state the number of novel possible exploits per category in Table~\ref{tab:stat-exploits}.

\begin{table}[]
    \centering
    \caption{Category-wise examples of novel possible exploits }
    \begin{tabular}{M{1.5cm}|M{6.25cm}}
     \textbf{Category} & \textbf{Novel possible exploit} \\
     \hline
     SDN-CP & Drop control messages to VNFs $\longrightarrow$ Disconnect targeted links in the network\\ \hline
     SDN-DP & Gain control of tenant controller VM $\longrightarrow$ Hijack network policy database\\ \hline
     NFV & Install a malicious switch $\longrightarrow$ Modify critical files on virtual storage or virtual compute\\ \hline
     SDN-CP, SDN-DP  & Hijack northbound API $\longrightarrow$ Input invalid data to tenant controller, forcing it to go to an invalid state\\ \hline
     NFV, SDN-CP & Exploit backdoors for testing $\longrightarrow$ Poison tenant controller host profile reservoir\\ \hline
     SDN-DP, NFV & Flood the switch table of target virtual switch $\longrightarrow$ Exploit the insecure crash recovery of NFV to shut down new VNFs assigned the same memory as the crashed VNF \\ \hline
     Malicious peripheral, SDN-DP & Compromise NFV-MANO unit $\longrightarrow$ Issue system command to terminate controller\\ \hline
     Malicious peripheral, NFV & Connect malicious peripheral and exploit MSI vulnerabilities $\longrightarrow$ Gain hypervisor privilege\\
    \end{tabular}
    \label{tab:exa-exploits}
\end{table}

\begin{table}[]
    \centering
    \caption{Number of novel possible exploits per category}
    \begin{tabular}{c|c}
     \textbf{Category} & \textbf{Number of novel possible exploits} \\
     \hline
     SDN-CP & 36\\
     SDN-DP & 23\\
     NFV & 36\\
     Malicious peripheral & 0\\
     Inter-graph & 24\\
     \hline
     \textbf{Total} & \textbf{119}\\ 
    \end{tabular}
    \label{tab:stat-exploits}
\end{table}

\subsection{ML Analysis}
\label{sec:Meth_ML_DAG}
When the number of components in the 5GCN increases, the size of the attack graphs increases 
significantly. To add a new node to these graphs, every possible connection between the new node and 
the existing nodes has to be analyzed manually. This is a tedious process that hinders scalability of 
this framework. To overcome this obstacle, we employ ML and CSP formulation to predict the 
possible connections of a new node in the graphs. 

\subsubsection{Feature Engineering}
Feature engineering is a necessary pre-processing step for using an ML or CSP model. Every 
possible branch in the graphs has to be represented by a feature vector for it to be processed 
by the ML or CSP model. We generate the feature vectors of a branch by implementing the 
following sequence of steps:

\begin{enumerate}
    \item Assign feature values for individual nodes.
    \item Combine the feature vectors of the constituent nodes of a branch.
\end{enumerate}

We assign various attributes (features) to the nodes of the attack graph(s) depending on the 
layer(s) at which it is executed, the type of impact the attack would have on the system and 
network, and its position in the graph(s). The exhaustive set of features that we used comprises 
the following: application layer, controller, application-controller interface, VNF, network 
infrastructure, management layer, hypervisor, flooding (DoS), access control, data plane, 
side-channel analysis (SCA), control channel, 
sensitive information, SDN-CP, SDN-DP, NFV, malicious peripheral, head, and tail. We assign $1$ 
to the features that are related to the node and $0$ to the others. For example, we demonstrate 
the features of nodes \textit{'Install malicious apps'} and \textit{'Assign low memory to VM'} 
in Table~\ref{tab:features}. We can observe that the feature 
vectors of these two nodes are 
$\{1,0,0,0,0,0,0,0,0,0,0,0,0,1,1,0,0,1,0\}$ and $\{0,0,0,1,1,1,1,1,0,0,0,0,0,0,0,1,0,0,0\}$.

\begin{table}[]
    \centering
    \caption{Node features}
    \begin{tabular}{c|M{1.5cm}|M{1.5cm}}
     \textbf{Feature} & \textbf{Install malicious apps} & \textbf{Assign low memory to VM}\\
     \hline
     Application layer & 1 & 0\\
     Controller & 0 & 0\\
     Application-controller interface & 0 & 0\\
     VNF & 0 & 1\\
     Network infrastructure & 0 & 1\\
     Management layer & 0 & 1\\
     Hypervisor & 0 & 1\\
     Flooding & 0 & 1\\
     Access control & 0 & 0\\
     Data plane & 0 & 0\\
     SCA & 0 & 0\\
     Control channel & 0 & 0\\
     Sensitive information & 0 & 0\\
     SDN-CP & 1 & 0\\
     SDN-DP & 1 & 0\\
     NFV & 0 & 1\\
     Malicious peripheral & 0 & 0\\
     Head & 1 & 0\\
     Tail & 0 & 0\\
    \end{tabular}
    \label{tab:features}
\end{table}

\begin{figure}[h]
\centerline{\includegraphics[scale=0.5]{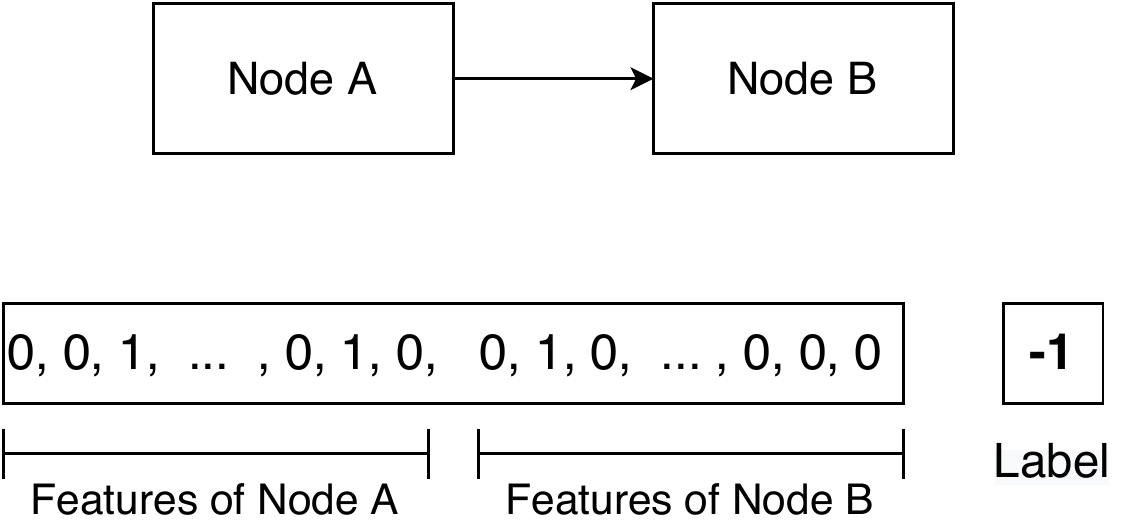}}
\caption{Constructing a feature vector and label for a plausible branch}
\label{fig:datapoint}
\end{figure}

We represent a branch of the graph with an ordered pair of the source and destination nodes, 
i.e., (source, destination). We obtain the feature vector of a branch by the ordered 
concatenation of the feature vectors of the source and destination nodes, as shown in 
Fig.~\ref{fig:datapoint}. This feature vector constitutes a datapoint for our ML/CSP model. We 
assign a positive label (equal to $1$) or a negative label (equal to $-1$) to this datapoint if 
the branch is feasible or infeasible, respectively.

We classify all plausible branches into positive and negative examples. The positive examples 
also include the existing branches in the attack graphs. We split the dataset for each graph into a 
training set and a test set.  The training set is used to train the model and the test set is 
used to evaluate it. The training set has 85\% of the data 
while the test set has the remaining. Table~\ref{tab:stat-examples} shows the number of instances in the training and test sets for each of the graphs.

\begin{table}[h!]
    \centering
    \caption{Number of datapoints per graph}
    \begin{tabular}{c|c|c}
     \textbf{Graph} & \textbf{Training set} & \textbf{Test set} \\
     \hline
     SDN-CP & 552 & 98\\
     SDN-DP & 898 & 159\\
     NFV & 510 & 90\\
     Malicious peripherals & 690 & 122\\
     Inter-graph connections & 6548 & 1156\\
     \hline
     \textbf{Total} & \textbf{9198} & \textbf{1625}\\ 
    \end{tabular}
    \label{tab:stat-examples}
\end{table}

\subsubsection{Analysis with a CSP Formulation}
\label{sec:csp}
A CSP formulation requires creating a set of constraints on the features of the data instances, such that any feature that satisfies all the constraints represents a feasible exploit.

To obtain a CSP formulation based on our dataset, we generate a set $S$ that contains the
feature vectors of all the positive examples in the training set. For prediction, 
we check if the feature vector of the test instance belongs to $S$. If it does, we assign a positive 
label to it; else, a negative one. 
%For example, let the set of features of all positive examples in the training set be 
%$S = \{[1,1,1,1], [1,1,0,0], [1,0,0,1]\}$. Let two test instances be $t_1=[1,1,1,1]$ and 
%$t_2 = [0,0,0,0]$. According to our CSP formulation, we assign a positive label to $t_1$ because 
%it belongs to $S$. We assign a negative label to $t_2$ because it does not belong to $S$.

\subsubsection{ML models}
We train multiple ML models on our data and choose the best-performing ones for our 
final ensemble model. The performance of these models is reported in 
Section~\ref{sec:ML-perf-results}. In this section, we briefly describe the various ML models that 
we experiment with.

\begin{itemize}
\item \textbf{Naive Bayes:} Naive Bayes is a probabilistic ML algorithm based on Bayes theorem. 
The Naive Bayes model assumes that features are independent of each other, given the label.
%and every feature has an equal influence on the label. 
%These assumptions are not valid in many real-world scenarios, thus limiting the performance of this 
%algorithm. 
Let the class label be denoted by $y$ and the input features by $\{x_1, x_2, ..., x_{n}\}$. Assuming 
feature independence, the probability of label $y$ can be calculated as

\begin{equation}
\centering
P(y|x_1,x_2,...,x_n) = \frac{P(y) \prod_{i=1}^{n} P(x_i|y)}{\prod_{i=1}^{n}P(x_i)}
\end{equation}

The class label with the highest conditional probability is assigned to a test instance, i.e., $prediction = \argmax \limits_{y} P(y|x_1,...,x_{n})$. The probabilities $P(x_i|y)$, $P(x_i)$, and $P(y)$ can be obtained by constructing a frequency table of the features from the training data. In our experiments, we used the Gaussian Naive Bayes classifier, where the likelihood of the features is assumed to be a Gaussian distribution.

\item \textbf{Decision Tree:} A decision tree classifier uses a decision tree to assign class labels. 
A decision tree can be expressed as a logical expression composed of 'AND' and 'OR' boolean operators. 
The leaf nodes of the tree represent the class labels. The other nodes represent conditional tests on 
the data attributes. Edges between two nodes represent control flow transition that depends on the 
outcome of the conditional test at the source node. 
%An example of a decision tree is shown in Fig.~\ref{fig:DT}.

%\begin{figure}[h]
%\centerline{\includegraphics[scale=0.5]{pics/DT.pdf}}
%\caption{Decision tree for determining the broad category of Olympic men's boxing championship}
%\label{fig:DT}
%\end{figure}

%The decision trees are constructed using divide and conquer algorithms. These models are fast to 
%execute, easy to interpret and useful in eliminating unimportant features. However, they are prone to 
%overfitting and can make counter-intuitive decisions when the sizes of the trees increase.

\item \textbf{k-Nearest Neighbors (k-NN):} The k-NN algorithm assigns a datapoint to the most popular 
class label among its $k$ ($k\geq1$) nearest neighbors. 
%This algorithm is very sensitive to the local structure of the data.
We experimented with $k = \{1,2,3,4,5,6,7\}$. We observed that the performance on our data initially 
increased with an increase in $k$ till $k=3$. Then, it either stopped increasing or started 
decreasing. Hence, we chose $k=3$.

\item \textbf{Support Vector Machine (SVM):} For an $n$-dimensional dataset, SVM constructs
an $(n-1)$-dimensional separating hyperplane that serves as the decision boundary. SVM can generate 
nonlinear decision boundaries with the help of kernel transformations based on a quadratic
optimization algorithm.
%An example of a non-linear decision boundary is shown in Fig.~\ref{fig:SVM}. 
%Unlike many contemporary ML algorithms, like k-NN, SVM does not use a greedy approach to compute the 
%decision boundary.  It uses a quadratic optimization algorithm, thus taking the global structure of 
%the data into consideration. A limitation of SVM is that the kernel function needs to be specified 
%and cannot be learnt. 

%\begin{figure}[h]
%\centerline{\includegraphics[scale=0.5]{pics/SVM.pdf}}
%\caption{SVM decision boundary on a 2-dimensional dataset}
%\label{fig:SVM}
%\end{figure}

We experimented with various parameters of the SVM model. We observed that the most effective kernel 
for our data was the \textit{radial basis function}. The class imbalance effects are mitigated 
through data preprocessing (see Section \ref{sec:ML-perf-results} for details). 
%We experimentally determined that a polynomial of degree $3$ yielded the most optimal results on 
%our data.

\item \textbf{Artificial Neural Network (ANN):} ANNs are loosely modeled after the biological neurons 
in the brain. We use an ANN variant called the multi-layer perceptron (MLP). 
%Fig.~\ref{fig:ANN} shows an example of an MLP architecture. 
The neurons in a MLP model are arranged in multiple layers. Every neuron receives signals from all the 
neurons in its previous layer. All these signals are weighted by their corresponding edge weights and 
their weighted sum is passed through a nonlinear activation function. This output is then propagated 
to all the neurons in the next layer. The training process involves updating the edge weights so that 
the prediction error is minimized.

%\begin{figure}[h]
%\centerline{\includegraphics[scale=0.5]{pics/ANN.pdf}}
%\caption{An example of an MLP architecture}
%\label{fig:ANN}
%\end{figure}

In our experiments, two-layer MLPs with the ReLU activation function yield the 
best results.

\end{itemize}

\subsubsection{Performance Results}
\label{sec:ML-perf-results}
In this section, we compare the performance of various ML algorithms on our data. 
We use the Negative Predictive Value (NPV) to evaluate the ML models. NPV is defined as the
fraction of correct negative predictions, as shown in the equation below.  We combine the models with 
highest NPVs to obtain our final ensemble model.  We evaluate the ensemble model with additional 
metrics like precision, recall, F1 score, Matthew's correlation coefficient (MCC), and classification 
accuracy.

\begin{equation}
    NPV = \frac{\text{True Negative}}{\text{True Negative} + \text{False Negative}}
\end{equation}

We design our framework in such a way that the security analyst, who uses our 
framework, can trust the negative predictions of our model with high confidence to be infeasible 
exploits. Then, the analyst only needs to manually examine the positive predictions for possible 
exploits. This significantly reduces the amount of manual effort needed. 
The NPVs of our models are shown in Table \ref{tab:ML-perf}.

We use stratified three-fold cross validation for evaluation of our models. 
%K-fold cross validation is effective in minimizing bias and overfitting of the ML model to the 
%training data. The process involves splitting the dataset into $k$ groups or folds. Then, we choose 
%one fold as the test set, fit the ML model on the remaining folds and evaluate the trained model on 
%the test fold. This process is repeated till each fold serves as the test fold, and the results are 
%then summarized. 
Stratified cross validation ensures that each fold has an equal ratio of positive and negative labels.

Our dataset is quite imbalanced with a much higher fraction of negative examples. 
To mitigate its impact, we resample the positive examples $n$ times, where the value of $n$ changes 
for different algorithms. The value of $n$ varies between $3$ and $12$. We observe that this is 
highly effective for all the ML models, except k-NN.

\begin{table}[h]
    \centering
    \caption{NPV (in \%) of ML/CSP models}
    \begin{tabular}{M{1.25cm}|c|c|c|M{1.4cm}|M{1.2cm}}
     \textbf{Algorithm} & \textbf{SDN-CP} & \textbf{SDN-DP} & \textbf{NFV} & \textbf{Malicious peripherals} & \textbf{Inter-graph}\\
     \hline
     Naive Bayes & 94.44 & 94.91 & 89.8 & 97.44 & \textbf{100}\\ \hline
     Decision tree & 90.7 & 94.63 & 89.02 & 98.26 & 99.74\\ \hline
     k-NN (k=3) & 87.76 & 93.67 & 87.64 & 97.54 & \textbf{100}\\ \hline
     SVM & 93.33 & 95.65 & 91.38 & 98.86 & 99.74\\ \hline
     Neural network & \textbf{95.0} & \textbf{96.9} & 91.67 & \textbf{98.98} & 99.82\\ \hline
     CSP & 94.38 & 96.69 & \textbf{95.12} & 98.35 & 99.65\\
    \end{tabular}
    \label{tab:ML-perf}
\end{table}

We select the models with the highest NPVs in Table~\ref{tab:ML-perf} and combine 
them into an ensemble model. For the inter-graph dataset, although the Naive Bayes and k-NN ($k=3$) 
models have perfect NPV values, we do not select these models. This is because Naive Bayes and 
k-NN have very low precision values of  $0.005$ and $0.008$, respectively, on this dataset. This 
overshadows their perfect NPV scores. The final ensemble model is shown in 
Table~\ref{tab:ensemble}. The numbers in the parentheses indicate the number of neurons in the
two hidden MLP layers.
%It is observed that models with a high NPV have low precision, recall and F1-scores, and vice versa. 
%Since, we prioritize NPV for our framework, our F1-scores are low.}

\begin{table}[h]
    \centering
    \caption{The final ensemble model} 
%MLP $(a,b)$ denotes a MLP having two layers of neurons, with the number of neurons in the first 
%and second layer being $a$ and $b$ respectively.}
    \begin{tabular}{c|c}
     \textbf{Graph} & \textbf{Algorithm}\\
     \hline
     SDN-CP & MLP (6,2)\\
     SDN-DP & MLP (8,2)\\
     NFV & CSP\\
     Malicious peripherals & MLP (5,2)\\
     Inter-graph & MLP (5,2)\\
    \end{tabular}
    \label{tab:ensemble}
\end{table}

In Table~\ref{tab:confusion-matrix}, we show the confusion matrix of the final ensemble model shown in 
Table~\ref{tab:ensemble}. The confusion matrix reports the true positives (TP), false positives
(FP), false negatives (FN), and true negatives (TN).

\begin {table}[h!]
\begin{center}
\caption {Confusion matrix of final ensemble model on the test set}
\begin{tabular}{c|c|c|c}
%\hline
 & \textbf{Actual = True} & \textbf{Actual = False} & \\
\hline
\textbf{Predicted = True} & 26 & 67 & 93\\

\textbf{Predicted = False} & 15 & 1517 & 1532\\
\hline
 & 41 & 1584 & 1625\\
\hline
\end{tabular}
\label{tab:confusion-matrix}		
\end{center}
\end {table}

In Table~\ref{tab:confusion-matrix}, we observe that there are 93 positive
predictions. Our framework reduces the search space of manual analysis to 93 instances from the
original search space of 1625 instances. This is a 94.3\% reduction in manual effort. Manual 
examination of these 93 instances leads to the discovery of the 26 true positives as novel possible 
exploits. The drawback of using the ML/CSP approximation is that we fail to include the 15 false 
negatives in our search space, thus missing the detection of 15 novel possible exploits.

We evaluate our final ensemble model with the following metrics:

\begin{itemize}
    \item \textbf{Precision:}  Precision is defined as
    
    \begin{equation}
    Precision = \frac{\text{TP}}{\text{TP} + \text{FP}}
    \end{equation}
    
%Precision returns a value between 0 and 1. The security analyst using our framework is expected to 
%manually examine only the positive predictions for possible exploits. 
A higher precision implies a lower FP. This implies that smaller manual effort is devoted to manually 
examining infeasible exploits, thus resulting in higher automation efficiency.
    
    \item \textbf{Recall:} Recall of a model is defined as
    
    \begin{equation}
    Recall = \frac{\text{TP}}{\text{TP} + \text{FN}}
    \end{equation}
    
%It returns a value between 0 and 1. 
A high recall value enables the user of our framework to discard the negative predictions from the 
manual examination set with high confidence. This requires our model to have a minimal FN.
    
    \item \textbf{F1 score:} F1 score is the harmonic mean of precision and recall and is defined as
    
    \begin{equation}
    F1 = 2 \times \frac{Precision \times Recall}{Precision + Recall}
    \end{equation}
    
    F1 score aims to strike a balance between precision and recall. It is a useful metric when there 
is an uneven class distribution. 
%It returns a value between 0 and 1.
    
    \item \textbf{MCC:} MCC is a measure of quality of a binary prediction algorithm. It performs 
well even for imbalanced classes. It returns a value between $-1$ and $+1$. $-1$ corresponds to a 
complete disagreement between observation and prediction, $0$ corresponds to random guessing, 
and $+1$ corresponds to a perfect prediction system. It is defined as
    
    \begin{equation}
    MCC = \frac{\text{TP} \times \text{TN} - \text{FP} \times \text{FN}}{\sqrt{\text{(TP+FP)(TP+FN)(TN+FP)(TN+FN)}}}
    \end{equation}
    
    \item \textbf{Accuracy:} Accuracy portrays the overall performance of the framework. It is defined as
    
    \begin{equation}
    Accuracy = \frac{\text{TP} + \text{TN}}{\text{TP + FP + TN + FN}}
    \end{equation}
    
\end{itemize}

We present the various performance metrics of our final ensemble model in 
Table~\ref{tab:performance-ensemble}. We obtain a high NPV and accuracy. However, our precision, 
recall and F1 scores are not as impressive as the NPV and accuracy. Our experiments show that there 
is a trade-off among the various metrics. If we choose to construct our final ensemble model with 
a high F1 score, then the NPV suffers. This is a trade-off that has to be made by the security 
analyst. Since we prioritize NPV, our precision and F1 scores suffer.

\begin {table}[h!]
\begin{center}
\caption {Performance metrics of final ensemble model on the test set}
\begin{tabular}{c|c}
\hline
\textbf{Metric} & \textbf{Value} \\
\hline
NPV & 0.99 \\

Precision & 0.28 \\

Recall & 0.63 \\

F1 score & 0.36 \\

MCC & 0.4 \\

Accuracy & 0.95 \\

\end{tabular}
\label{tab:performance-ensemble}		
\end{center}
\end {table}

\section{Case Study I: 5G-AKA}
\label{sec:5GAKA}
The novel possible exploits of the 5GCN reported in the previous sections can lead 
to novel attacks at the higher layers of the network or increase the ease of execution of existing 
attacks in the protocol and application layers. In this section, we demonstrate the impact of 5GCN 
vulnerabilities on the protocol layer.

AKA is used in telecommunication networks to establish a secure and authenticated connection 
between the subscribers and service providers. It enables sharing of a secret key between the 
user and service provider that is used to secure all further communication.

The AKA protocols have evolved through generations of telecommunication networks. Today, the
most widely used authentication mechanism in such networks is the 4G-AKA. The 3GPP Consortium 
has designed 5G-AKA to provide superior privacy and security guarantees than 4G-AKA. However, 
it has been shown that multiple 4G-AKA vulnerabilities still persist in 
5G-AKA~\cite{koutsos20185g}. 5G-AKA is also vulnerable to novel attacks that were not possible 
in previous generations of networks~\cite{basin2018formal, cremers2019component, 
edris2020formal}. The 5G-AKA protocol can be easily compromised if the 5GCN is vulnerable. 
In this section, we analyze the implications of our novel  possible exploits  on 5G-AKA security.

\subsection{5G-AKA protocol}
The 5G-AKA protocol authenticates a user equipment (UE), a serving network (SN), and a home 
network (HN) to each other.  It is a challenge-response 
based protocol where the UE is authenticated as a legitimate user only if it succeeds in 
providing the expected response to a challenge provided by the HN. Unlike previous networks, 
the identity of the UE, called subscriber permanent identifier (SUPI) in 5G networks, is not 
sent directly. In 5G networks, the UE sends a subscriber concealed identifier 
(SUCI) that prevents international mobile subscriber identity catcher attacks~\cite{koutsos20185g}. 

\begin{figure}[t]
\centerline{\includegraphics[scale=0.5]{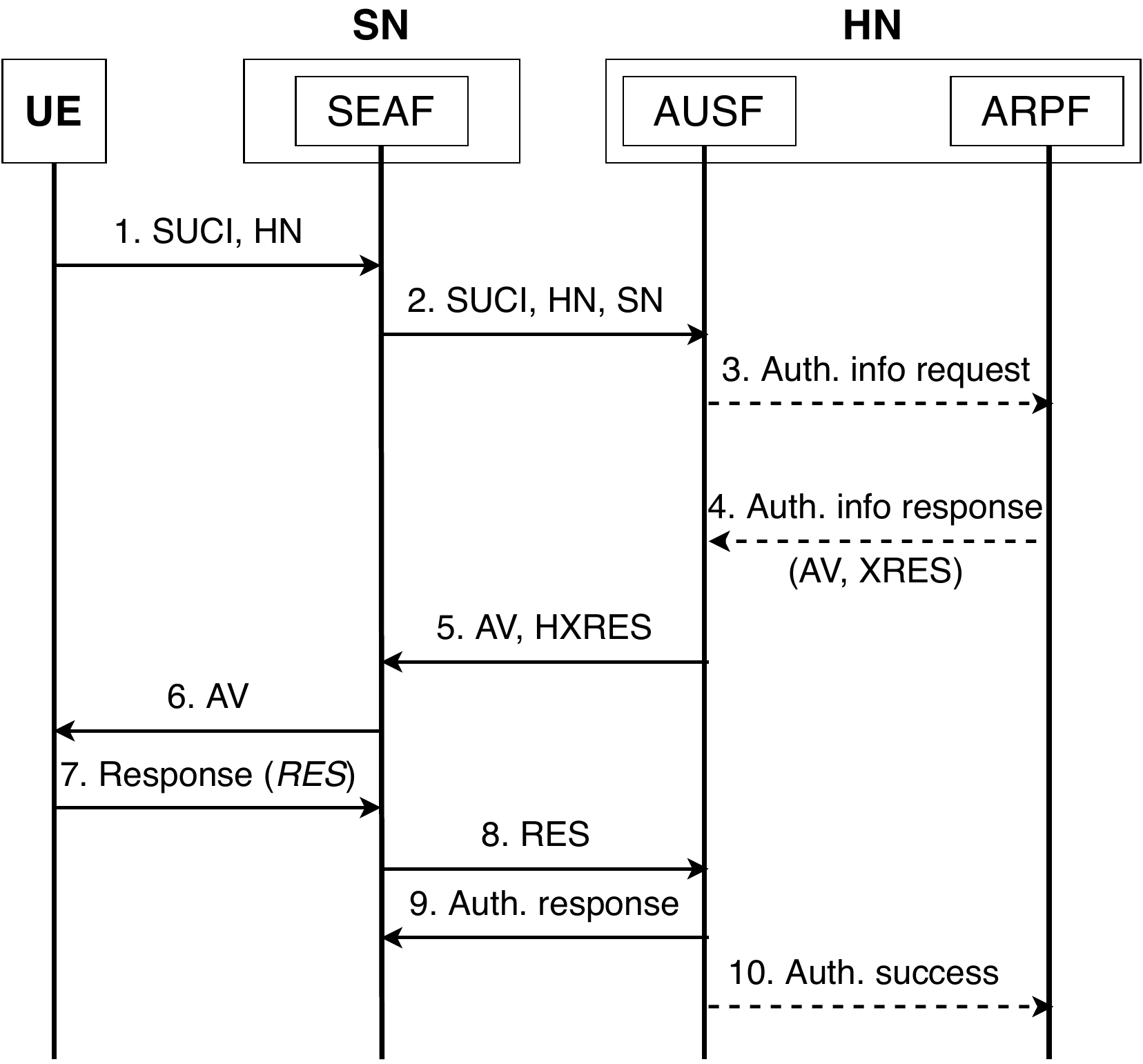}}
\caption{5G-AKA protocol flow. The dashed lines represent secure channels.}
\label{fig:5g-aka}
\end{figure}

Often, the SN and HN are the same network. However, sometimes they are different. For example, 
when a UE is roaming, its SN is different from its HN. In our analysis, we consider a separate 
SN and HN because this scenario is more prone to attacks. The primary network functions involved 
in 5G-AKA are the Authentication Server Function (AUSF), Authentication Credential Repository 
and Processing Function (ARPF), and Security Anchor Function (SEAF). A simplified outline of the 
5G-AKA protocol is shown in Fig.~\ref{fig:5g-aka}. The details of the messages are abstracted 
for simplicity.  \textit{AV} denotes the authentication vector, \textit{XRES} denotes the 
expected response from the UE, and \textit{HXRES} denotes a hash of XRES.

\subsection{Threat Model}
Most of the security analysis of the 5G-AKA protocol so far has considered a threat model where 
the adversary has access to the UE and the communication channels between different networks. The core network infrastructure is considered to be inaccessible to the adversary.
From Fig.~\ref{fig:5g-aka}, we see that the AUSF and ARPF communicate over a secure network 
because they belong to the same network (HN). However, we have shown in 
Section~\ref{sec:Methodology} that the internal components of the 5GCN can be compromised. To 
overturn the assumption of having an impenetrable 5GCN, 
we expand the attack surface of the 5G-AKA protocol in our analysis. In our threat model, an 
adversary can compromise the network's private channels and the network functions as well. In 
Section~\ref{sec:Methodology}, we demonstrated how VNFs and other network components can be 
compromised by exploiting vulnerabilities of SDN, NFV, and IO peripherals.

\subsection{5G-AKA Security Analysis}
In this section, we analyze the implications of a compromised 5GCN on the security 
properties of the 5G-AKA protocol. A compromised 5GCN leads to unique exploits and also 
facilitates exploits that were unrealistic before. Section~\ref{sec:5G-aka-attacks} describes 
the novel attack vectors that become possible using our analysis framework to compromise the 5GCN. Section~\ref{sec:5G-aka-property} analyzes the various 5G-AKA security properties that are 
violated in the presence of a compromised 5GCN.

\subsubsection{Novel Attacks}
\label{sec:5G-aka-attacks}
The vulnerabilities of NFV, SDN, and IO peripherals have a variety of potential consequences at 
the network level. These consequences include flooding (DoS) attacks, termination of sensitive 
VNFs, passive MiTM attacks (like eavesdropping), hijacking of VNFs, and active MiTM attacks 
(like modification of in-flight traffic). In this section, we analyze how these consequences 
can be exploited to compromise the 5G-AKA protocol.

\begin{itemize}
    \item \textbf{Flooding attacks}: The 5G-AKA protocol is vulnerable to session confusion 
attacks triggered by a race condition in the AUSF-ARPF channel~\cite{cremers2019component}. We 
demonstrate this attack in Fig.~\ref{fig:race-cond}. When the ARPF receives multiple 
authentication requests in parallel, it sends the AVs for all the requests to the AUSF at the 
same time. This leads to a race condition in which the AUSF is unable to distinguish which AV 
belongs to which UE. Thus, there is a high probability that the AUSF sends the wrong 
credentials to the users. This is a probabilistic attack whose success rate can be increased with more parallel 
authentication requests from the adversary.
    
    \begin{figure}[h]
    \centerline{\includegraphics[scale=0.5]{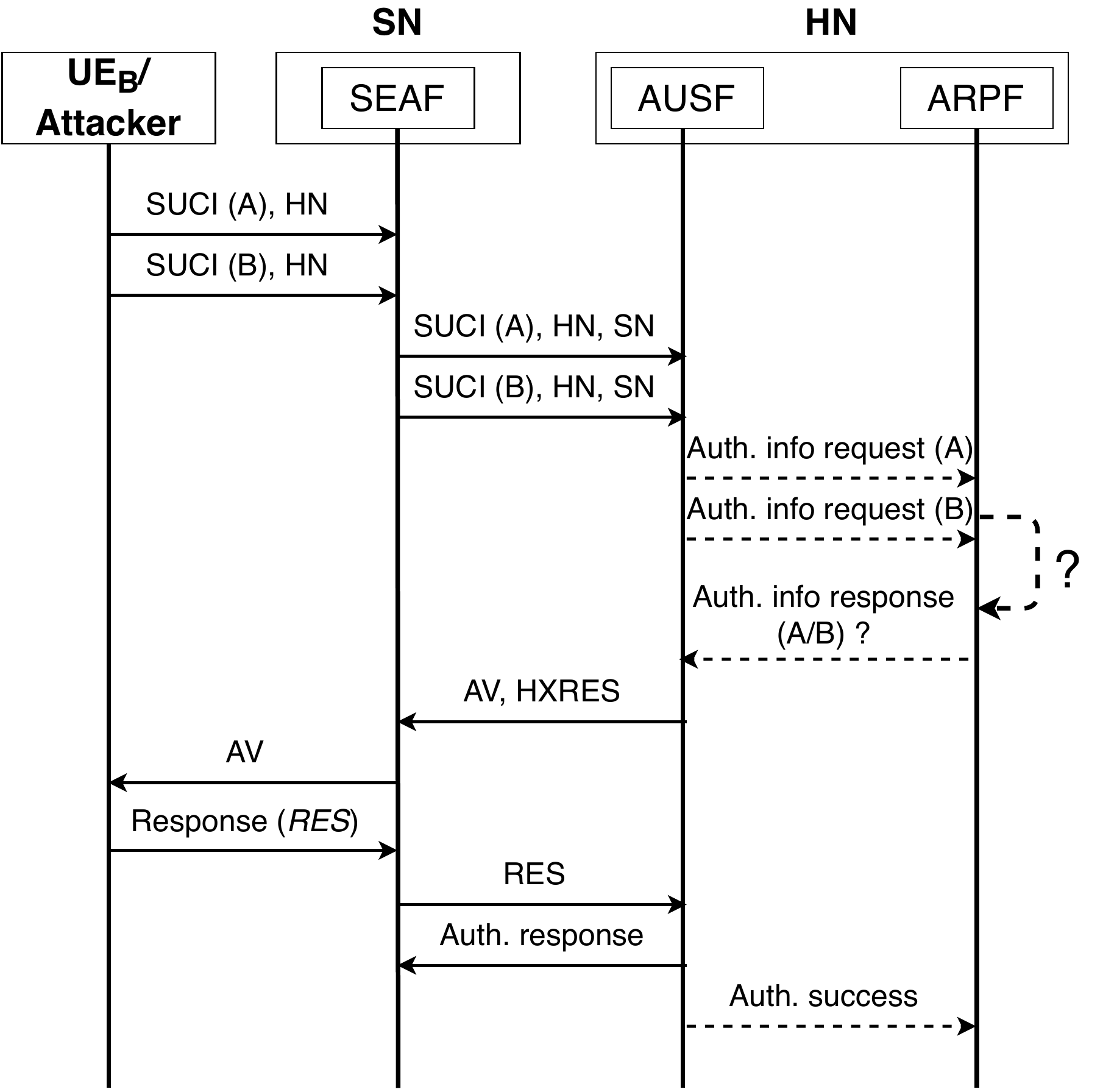}}
    \caption{The 5G-AKA session confusion attack flow}
    \label{fig:race-cond}
    \end{figure}

In the threat model of this attack, the adversary can hijack the VNFs on the
SN but not the VNFs on the HN. He can hijack the SEAF on the SN and use it to bombard the AUSF 
with multiple network packets of SUCI(Attkr) simultaneously. The AUSF generates authentication 
requests, \textit{Auth. info. request (Attkr)}, for all of these packets and sends them to the ARPF. 
When the ARPF receives all these packets simultaneously, along with \textit{Auth. info. request 
(Victim)}, it leads to a race condition. According to the 5G-AKA protocol specifications, the response 
of the ARPF does not include the identity of the UE. Thus, the simultaneous reception of multiple 
(AV, XRES) pairs by the AUSF causes a session confusion. It is probable that the AUSF forwards the 
AV of the victim to the adversary as a result of this confusion. Now, the adversary can authenticate 
himself as the victim.

The probability of success of this attack is $(1-\frac{1}{n})$, where $n$
depicts the total number of simultaneous authentication requests received by the ARPF. 
In Fig.~\ref{fig:race-cond}, the value of $n$ is $2$; thus the probability of attack success is 0.5. 
Increasing the number of simultaneous authentication requests from the adversary's UEs (by 
replay/flooding attacks from a compromised SEAF) will increase $n$, thus increasing the probability of attack success. The 
node ``Flood VNF with requests/high priority requests" of the NFV attack graph in 
Fig.~\ref{fig:nfv-threat} can be implemented via multiple possible exploits discovered by our 
framework to execute this attack.

\begin{comment}    
    \begin{figure}[h!]
    \centerline{\includegraphics[scale=0.55]{pics/prob.pdf}}
    \caption{Success probability of the session confusion attack}
    \label{fig:prob}
    \end{figure}
\end{comment}
    \item \textbf{Termination of sensitive VNFs:} NFV and IO vulnerabilities can be exploited to 
forcibly terminate targeted VNFs. This can be achieved by executing one of the 
following nodes in Fig.~\ref{fig:nfv-threat}: ``\textit{Shutdown sensitive VNFs}," 
``\textit{VNF crash}," ``\textit{DoS on target VNF}." Our analysis framework predicts multiple possible exploits for implementing these nodes in a vulnerable 5GCN. Untimely 
termination of SEAF, AUSF or ARPF disrupts the 
5G-AKA protocol. Although the adverse effects of such attacks can be mitigated by a 
fault-tolerant implementation of these functions~\cite{leu2020fault}, all ongoing 
authentication information is lost. This forces the UEs to restart the 5G-AKA protocol.
    
    \item \textbf{Passive MiTM}: Passive MiTM can be executed on the AUSF-ARPF channel. Since 
this channel is considered to be secure by the 5G-AKA designers, it is not required to be 
encrypted.  Operators would also prefer having no encryption to boost performance. In our 
analysis of attack graphs in Section~\ref{sec:Methodology}, we generated multiple attack 
vectors for launching privilege escalation attacks that give access to 5GCN resources. An 
adversary with access to the 5GCN infrastructure can eavesdrop on the secure channels. This 
leads to the disclosure of private information like AV, XRES, SUPI, and the secret keys of 
AUSF and SEAF to the adversary.  The adversary can exploit the knowledge of XRES and SUPI to 
authenticate himself on behalf of a legitimate UE. The secret key of AUSF can be exploited to 
authenticate a fake base-station, thereby launching active MiTM attacks on UEs.
    
    \item \textbf{Hijacking of VNFs}: Hijacking of sensitive VNFs like the SEAF, AUSF or ARPF 
can cause the 5G-AKA protocol to prevent authentication of legitimate UEs or authenticate 
adversaries with the credentials of a legitimate UE. Our methodology in 
Section~\ref{sec:Methodology} demonstrates multiple access control and privilege escalation 
attacks in the graphs that can be exploited to hijack VNFs.
    
    \item \textbf{Active MiTM}: Active MiTM attacks involve modifying the packets during 
transit.  This compromises the integrity of network packets. Since the connections in the same 
network are assumed to be secure in the original 5G-AKA threat model, the operators are not 
required to have integrity checks on intra-network messages. The adversary can get access to 
the internal network by exploiting certain infrastructure vulnerabilities and modify the packets 
in transit. Our methodology in Section~\ref{sec:Methodology} demonstrates multiple attack 
vectors for launching active MiTM attacks. The attack graphs in 
Fig.~\ref{fig:sdn-cp}, \ref{fig:sdn-dp}, and \ref{fig:mal-per} demonstrate that there are multiple 
openings for MiTM attacks in a vulnerable 5GCN. The adversary can exploit them to modify the AV, 
XRES, AUSF secret key or SUPI in the AUSF-ARPF channel without being detected. Modifying the SUPI or 
XRES will enable the adversary to authenticate himself on behalf of a legitimate UE. Modifying 
the AUSF secret key enables the user to launch a fake base station.
\end{itemize}

\subsubsection{5G-AKA Property Violations}
\label{sec:5G-aka-property}
The 3GPP Consortium has detailed the security requirements of 5G system components in TS 33.501 
v0.7.0~\cite{3gpp-aka}. The security requirements that are related to the 5G-AKA protocol 
can be expressed concisely through two secrecy properties and seven authentication 
properties~\cite{cremers2019component}. Every vulnerability of the 5G-AKA protocol, including 
the ones mentioned in Section~\ref{sec:5G-aka-attacks}, violates at least one of these security 
properties. Hence, analyzing these properties may provide insights into what kinds of attacks 
are possible. 

The secrecy properties of 5G-AKA are:

\begin{itemize}
    \item[] \textbf{S1}. The long-term secret key of the UE should be unknown to the adversary.
    \item[] \textbf{S2}. The adversary should not have access to the secret keys of AUSF and 
SEAF.
\end{itemize}

The authentication properties of 5G-AKA are:

\begin{itemize}
    \item[] \textbf{A1}. SN and UE must agree on the identity of UE.
    \item[] \textbf{A2}. UE and SN must agree on the identity of SN.
    \item[] \textbf{A3}. HN and SN must agree on the identity of UE.
    \item[] \textbf{A4}. UE and HN must agree on the identity of HN.
    \item[] \textbf{A5}. UE and HN must agree on the identity of SN.
    \item[] \textbf{A6}. UE, HN, and SN must agree on the anchor key of SEAF, $K_{SEAF}$.
    \item[] \textbf{A7}. UE, HN, and SN must agree that an anchor key $K_{SEAF}$ instance is not used more than once.
\end{itemize}

The security of the 5G-AKA protocol is compromised if any of the aforementioned properties is 
violated. It has been shown that the compromise of participating components of the 5G-AKA 
protocol leads to the violation of these properties~\cite{cremers2019component}. We demonstrate 
the consequences of compromising the 5GCN on the 5G-AKA properties in 
Table~\ref{tab:5G-AKA-property}.

\begin {table}[h!]
\begin{center}
\caption {Property satisfaction under compromised channels and components}
\begin{tabular}{M{3cm}|M{0.19cm}|M{0.19cm}|M{0.19cm}|M{0.19cm}|M{0.19cm}|M{0.19cm}|M{0.19cm}|M{0.19cm}|M{0.19cm}}
\hline
\textbf{Compromised element} & \textbf{S1} & \textbf{S2} & \textbf{A1} & \textbf{A2} & \textbf{A3} & \textbf{A4} & \textbf{A5} & \textbf{A6} & \textbf{A7} \\
\hline
AUSF-ARPF channel; passive MiTM & \cmark & \xmark & \xmark & \xmark & \cmark & \cmark & \xmark & \xmark & \xmark\\ \hline
AUSF-ARPF channel; active MiTM & \cmark & \xmark & \xmark & \xmark & \cmark & \xmark & \xmark & \xmark & \xmark\\ \hline
SEAF & \cmark & \xmark & \xmark & \xmark & \xmark & \cmark & \xmark & \xmark & \xmark\\ \hline
AUSF & \cmark & \xmark & \xmark & \xmark & \xmark & \cmark & \xmark & \xmark & \xmark\\ \hline
ARPF & \cmark & \xmark & \xmark & \xmark & \cmark & \xmark & \xmark & \xmark & \xmark\\

\end{tabular}
\label{tab:5G-AKA-property}		
\end{center}
\end {table}

We see that 5GCN vulnerabilities and threats lead to the violation of many of the security 
properties of the 5G-AKA protocol. This demonstrates that 5GCN vulnerabilities also make
the 5G-AKA protocol vulnerable.

\section{Case Study II: WhatsApp Security in 5G Networks}
\label{sec:WhatsApp}
In this section, we analyze how various existing and novel possible exploits of a vulnerable 5GCN can lead to targeted attacks in the application layer of the network. We chose the WhatsApp application for our security analysis.

WhatsApp is the most widely used IM application in the world, with over 
1.5 billion users~\cite{wa-stats}. It is also one of the most secure IM applications, where all 
communications are end-to-end (E2E) encrypted. In this section, we demonstrate that even 
WhatsApp can be compromised through network and protocol vulnerability exploits. Various WhatsApp attack vectors that are facilitated by our 
methodology include the following.

%WhatsApp communication can be segregated at the network level using advanced network traffic 
%classification methods~\cite{sengupta2019exploiting}. Active MiTM attacks can be launched on 
%this traffic at the network level. 

\begin{itemize}
    \item \textbf{Impersonation of the victim via 5G-AKA}: As described in 
Section~\ref{sec:5G-aka-attacks}, the adversary can authenticate himself as the victim during 
5G-AKA protocol execution by exploiting any of the following attacks: flooding, passive MiTM, 
hijacking of VNFs, and active MiTM. Then, the adversary can use the victim's identity to 
impersonate him on WhatsApp.
    
    \item \textbf{Assisting WhatsApp impersonation through voicemail cracking}: During 
registration of a WhatsApp account, the user can choose to be authenticated by a text message 
or a call. If the user chooses to be authenticated by a call and fails to receive the 
authentication call, then the one-time password voice message is 
saved in voicemail. It has been shown that voicemails can be easily hacked using brute-force 
attacks~\cite{wa-voicemail}.  This attack has a low probability of being successful in a 
real-world situation because it requires the victim to either be offline or ignore the 
authentication call. This obstacle for the adversary can be bypassed by launching a DoS 
attack on the victim's network infrastructure. The framework discussed in Section 
\ref{sec:Methodology} generates multiple novel possible exploits to launch a DoS attack on various 
components of the 5GCN. Fig.~\ref{fig:sdn-cp}, \ref{fig:sdn-dp}, \ref{fig:nfv-threat}, and 
\ref{fig:mal-per} show that DoS attacks can be launched on VNFs, VMs, switches, and SDN controllers.
A DoS attack on the network infrastructure will terminate the victim's connection to the 5GCN, thus 
ensuring that he is offline. Now, the voicemail attack has a much higher probability of being 
successful.
    
    \item \textbf{Compromising encryption keys}: E2E security of WhatsApp can be readily 
compromised if the adversary gets access to the WhatsApp encryption keys on the device. The 
WhatsApp keys are stored in a sandbox memory on the smartphone that is only accessible by the 
WhatsApp application.  If an adversary has root privileges on the phone, he can access the 
WhatsApp encryption keys.  Rootkits can be installed on the UE by combining MiTM attacks in 
our attack graphs with baseband attacks~\cite{xenakis2015attacking}. Attack 
vectors that exploit rootkit injection attacks are described in Fig.~\ref{fig:sdn-dp}.
    
    \item \textbf{Lack of certificate pinning}: WhatsApp does not implement certificate pinning 
on the UE~\cite{wa-cert-pin}. This makes the WhatsApp clients vulnerable to MiTM attacks through 
certificate proxying. We demonstrated the possible exploits for launching an 
MiTM attack at the network level in Section~\ref{sec:Methodology}.  These attacks can be %successfully 
executed in the absence of certificate pinning.
\end{itemize}

%\begin{comment}
\section{Discussion}
\label{sec:Discussion}
The attack graphs depicted in Fig.~\ref{fig:sdn-cp}--\ref{fig:mal-per} are designed 
to be as exhaustive as possible. We have attempted to include all possible attack classes applicable 
to SDN, NFV, and malicious peripherals in a 5GCN in these graphs. For application of our framework 
to a specific 5GCN implementation, we have to derive 5GCN-specific graphs from the generalized graphs 
that we have presented. For a given 5GCN architecture, the relevant nodes from the generalized graphs 
are extracted to form the architecture-specific graphs. For example, if a 5GCN does not use LLDP to 
establish network topology, we will eliminate the LLDP-specific nodes from Fig.~\ref{fig:sdn-cp} for 
this 5GCN. If a 5GCN has a feature that warrants addition of new nodes to the graphs, we can use 
ML to predict the connections of the new nodes to the existing nodes. Thus, we can add new nodes to 
the graphs and create a 5GCN-specific attack graph for further analysis.

Probabilistic attack graphs, more popularly known as Bayesian attack graphs, have 
been extensively used to assess the security risk of networks. The framework proposed here can be 
extended to Bayesian attack graphs with minimal modifications. In a traditional Bayesian attack 
graph, each node represents a state of the system. An edge from state A to state B  exists if an 
exploit of a vulnerability at state A takes the system to state B. The weight of this edge is equal 
to the probability of execution of the aforementioned exploit. Hence, the graphs presented in our 
article can be transformed into equivalent Bayesian attack graphs if the edges have weights 
corresponding to their probability of execution. These probabilities can be obtained for
specific systems from the CVE databases. However, our framework is more useful than Bayesian attack 
graphs because it can also discover novel possible exploits in a system.
%\end{comment}

\section{Conclusion}
\label{sec:Conclusion}
5G communication systems have a huge potential for revolutionizing the way we live. This is made 
possible by the integration of new technologies like NFV and SDN into the 5GCN. This
gives rise to new vulnerabilities in the 5G system. In this article, we analyzed how various 
vulnerabilities of NFV, SDN, and malicious IO peripherals can interact with each other to 
compromise the security of the 5GCN. We discovered 119 novel possible exploits by analyzing the 
underlying patterns in the 113 existing attack vectors in SDN, NFV, and IO peripherals. We 
showed that a compromised 5GCN may have devastating consequences on the end user. A compromised 
5GCN was shown to trigger five unique types of attacks in the 5G-AKA protocol. These attacks 
can be further combined with infrastructure vulnerabilities to compromise targeted users at the 
application layer. We demonstrated this by analyzing four potential security loopholes in the 
WhatsApp IM application.

%\section*{Acknowledgments}
%The authors would like to thank NSF for supporting this work under Grant CNS-1617628.

\bibliographystyle{IEEEtranN}
\bibliography{bibtex/bib/IEEEexample}

\begin{IEEEbiography}[{\includegraphics[width=1in,height=1.25in,clip,keepaspectratio]{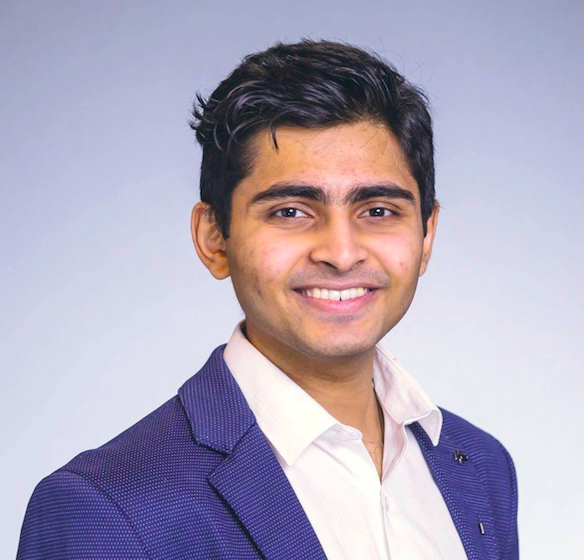}}]{Tanujay Saha}
Tanujay Saha is currently pursuing his Ph.D. degree at Princeton University, NJ, USA. He received his Master's Degree in Electrical Engineering from Princeton University and Bachelors in Technology in Electronics and Electrical Communications Engineering from Indian Institute of Technology, Kharagpur, India in 2017. He has held research positions in various organizations and institutes like Intel Corp., KU Leuven, and Indian Statistical Institute. His research interests lie at the intersection of IoT, cybersecurity, machine learning, embedded systems, and cryptography.
\end{IEEEbiography}

\begin{IEEEbiography}[{\includegraphics[width=1in,height=1.25in,clip,keepaspectratio]{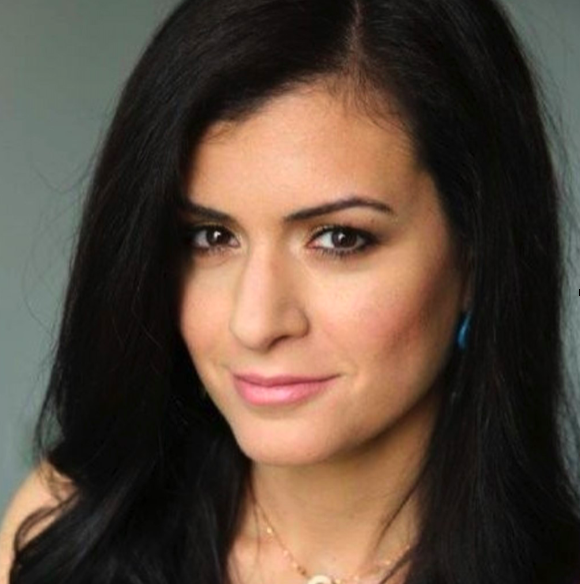}}]{Najwa Aaraj}
Najwa Aaraj is a Chief Research Officer at the UAE Technology
Innovation Institute. She holds a Ph.D. in Electrical Engineering from
Princeton University and a Bachelor’s in Computer and Communications
Engineering from the American University in Beirut. Her expertise lies
in applied cryptography, trusted platforms, secure embedded systems,
software exploit detection/prevention, and biometrics. She has over 15
years of experience working in the United States, Australia, Middle
East, Africa, and Asia with global firms. She has two patents and 15
academic publications. She has worked in a cybersecurity start-up
(DarkMatter). Prior to joining DarkMatter, she worked at Booz \& Company,
where she led consulting engagements in the communication and technology
industry for clients across four continents. She has also held research
positions at IBM T. J. Watson Center, New York, Intel Security Research
Group, Portland, Oregon, and NEC Laboratories, Princeton, New Jersey.
\end{IEEEbiography}

%\newpage

\begin{IEEEbiography}[{\includegraphics[width=1.25in,height=1.25in,clip,keepaspectratio]{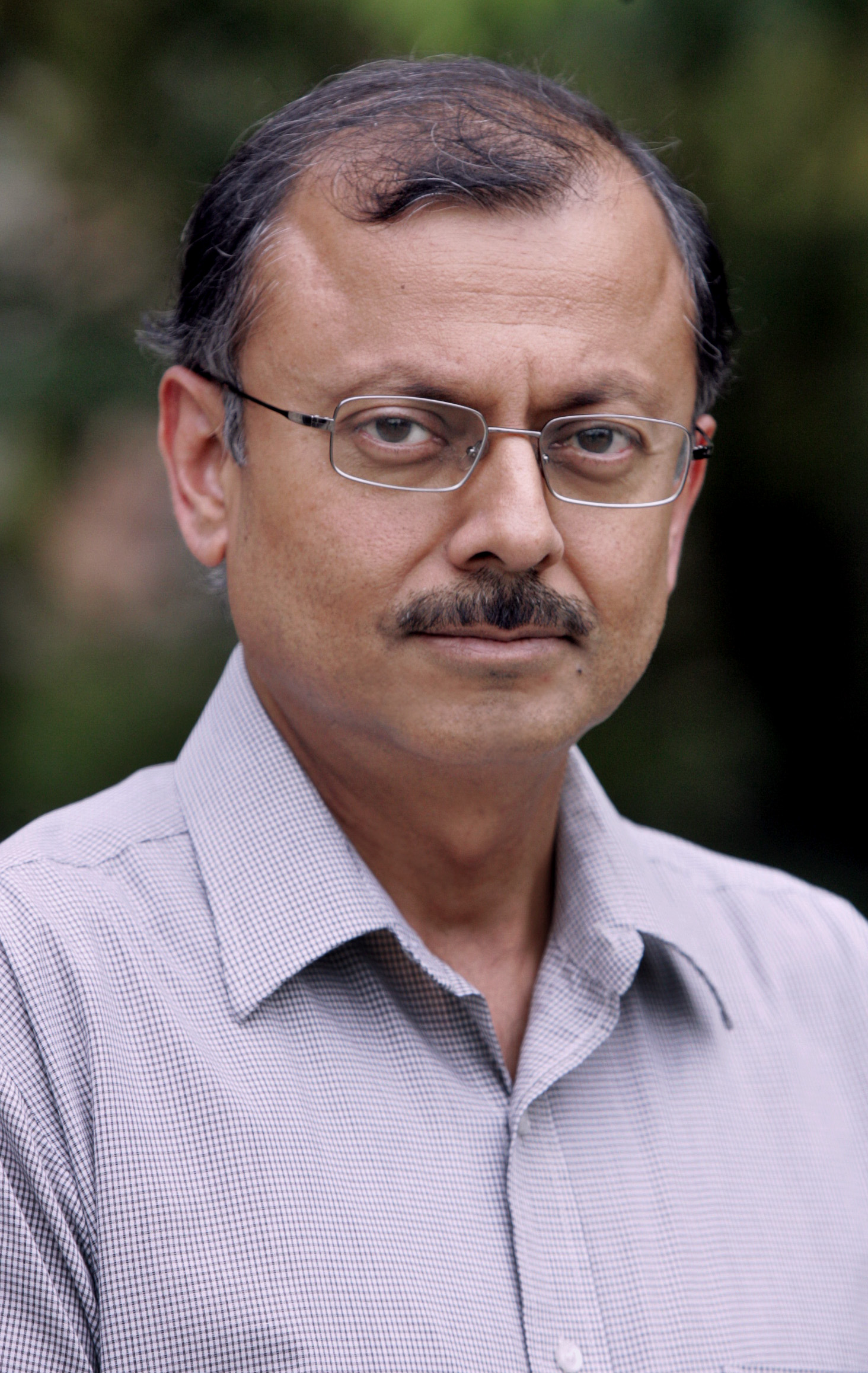}}]{Niraj K. Jha}
Niraj K. Jha received the B.Tech. degree in electronics and electrical 
communication engineering from I.I.T., Kharagpur, India, in 1981, and the 
Ph.D. degree in electrical engineering from the University of Illinois at
Urbana-Champaign, Illinois, in 1985. He has been a faculty member of the
Department of Electrical Engineering, Princeton University, since 1987.
He was given the Distinguished Alumnus Award by I.I.T., Kharagpur. He
has also received the Princeton Graduate Mentoring Award. He has served
as the editor-in-chief of the IEEE Transactions on VLSI Systems and as
an associate editor of several other journals. He has co-authored five
books that are widely used. His research has won 20 best paper awards or
nominations. His research interests include smart healthcare,
cybersecurity, machine learning, and monolithic 3D IC design. He has
given several keynote speeches in the area of nanoelectronic design/test
and smart healthcare. He is a fellow of the IEEE and ACM.
\end{IEEEbiography}

\end{document}